\documentclass[twocolumn,superscriptaddress,showpacs,notitlepage]{revtex4-1}
\usepackage{amsmath,amssymb}
\usepackage{graphicx}
\usepackage{txfonts}
\usepackage{color}
\usepackage{hyperref}
\usepackage{bbold}
\usepackage{framed}
\usepackage{wrapfig}
\usepackage[letterpaper]{geometry}
\global\long\def\ket#1{\left| #1\right\rangle }
\global\long\def\bra#1{\left\langle #1 \right|}
\global\long\def\av#1{\left\langle #1 \right\rangle }

\begin{document}

\title{Winding numbers of nodal points in Fe-based superconductors}

\author{Dmitry V. Chichinadze and Andrey V Chubukov}
\affiliation{School of Physics and Astronomy,
University of Minnesota, Minneapolis, MN 55455, USA}

\begin{abstract}
We analyze the nodal points in multi-orbital Fe-based superconductors from a topological perspective.   We consider the $s^{+-}$ gap structure with accidental nodes,  and the  $d$-wave gap  with  nodes along the symmetry directions.  In both cases, the nodal points  can be moved by varying an external parameter, e.g., a degree of inter-pocket pairing.   Eventually, the nodes merge and annihilate via a Lifshitz-type transition.  We discuss the Lifshitz transition in Fe-based superconductors from a topological point of view.  We show, both analytically and numerically, that the merging nodal points have winding numbers of opposite sign.  This is consistent with the general reasoning that the total winding number is a conserved quantity in the Lifshitz transition.
\end{abstract}

\maketitle

\section{Introduction}

The research on correlated electron systems over the last  decade have shown tremendous developments in two seemingly different areas.
 One area, generally termed as "topology in condensed matter", focuses on  topological description of quantum materials, with emphasis on specific invariants which characterize a particular quantum state of matter, and change only when a system undergoes a transition from one quantum state to the other. The research in this field started in the early 80's \cite{Thouless1982, Simon1983, Berry1984}, but rapidly accelerated over the last decade and led to qualitative new understanding of the properties of existing materials and to discoveries of numerous new materials exhibiting fundamentally novel properties \cite{Bernevig2006, Bernevig2006Science,Wan2011, Xu2015}.  Another area  is high-temperature superconductivity. The research in this field started after the discovery of SC in the cuprates and acquired a new dimension with the invent of  Fe-based superconductors (FeSCs) with multiple relevant orbitals and, as a consequence, multiple  Fermi pockets of hole and electron-type \cite{Kamihara2008}.

Some FeSCs exhibit superconducting properties consistent with the full gap, while the others show behavior consistent with gap zeros on some of the Fermi surfaces \cite{delaCruz2008, Liu2010, Putti2010}. A number of theories have been put forward about  $s$-wave superconductivity in FeSCs with  orbitally-induced  gap anisotropy.
When the anisotropy is strong enough, an $s$-wave gap can have nodes on some of the pockets~\cite{Hanaguri2010,Maier2011,Hirschfeld2011,Khodas2012,Chubukov2016}.  Because the gap nodes are accidental, they can appear or disappear via a Lifshitz-type transition~\cite{Lifshitz1960} under the change of external parameters like doping or pressure~\cite{Liu2010, Liu2011}.  In special cases (which we discuss below), the transition from a nodal state to a state with a full gap is more involved, with additional nodal points appearing near a transition and then annihilating the existing nodes~\cite{Hinojosa2015}.  Another set of theories for FeSCs analyzed  possible $d$-wave superconductivity, particularly in systems where only hole or only electron pockets are present.  In a one-band system a $d$-wave superconductor has symmetry protected gap nodes on the Fermi surface.  In multi-band materials, like FeSCs, these nodes can also be manipulated by, e.g., varying the strength of the inter-band pairing~\cite{Chubukov2016,Nica2017}. In the presence of such terms, the nodal points of the fermionic dispersion in a $d$-wave superconductor shift  from the original Fermi surfaces to the area between the pockets, come closer to each other  and eventually annihilate and disappear, leaving a $d$-wave superconductor with a full gap~\cite{Chubukov2016,Nica2017}.

  In this paper we discuss Lifshitz transitions in FeSCs from a topological viewpoint.  We argue that, while the symmetry of a superconducting state ($d-$wave or $s-$wave) does not change upon the apperance/disapperance of the zeros in the fermionic dispersion, the topological properties of a system do change because each nodal point is characterized by a particular winding number, which remains invariant as long as a nodal point exists, but vanishes once it disappears.

 We study two models of FeSCs,  one with an $\mathit{s}$-wave gap symmetry and accidental gap nodes, and another with a $\mathit{d}$-wave gap and nodes along particular symmetry directions.  In both models, the nodes can be manipulated by changing one or more model parameters. As a result, a system may undergo a  Lifshitz transition in which the nodal points merge and disappear.  We show, both analytically and numerically, that the merging nodal points have opposite sign winding numbers.  This is consistent with the general reasoning that the total winding number is a conserved quantity in the Lifshitz transition. We also show that when a pair of nodal points is spontaneously generated by changing an external parameter, the winding numbers of the two emerging nodes are opposite.

The merging and annihilation of nodal points  has been well studied  in Dirac and Weyl semi-metals,  which undergo a transition into an insulator under a variation of certain system parameters~\cite{Vafek2014}. Several authors have shown that in a semi-metal-to-insulator transition, the merging nodal points have opposite winding numbers \cite{Murakami2008,Murakami2007}.
 We demonstrate here that the same is true in nodal-to-full gap transitions in $s$-wave and $d$-wave superconductors.

The structure of the paper is as follows. In Sec. \ref{sec_2} we introduce the Berry curvature $B_{\vec{k}}$ and express the winding number as a particular 2D integral of $B_{\vec{k}}$. In Sec.  \ref{sec_3} we consider a model of an $s$-wave superconductor which undergoes a Lifshitz transition upon varying  one or more system parameters~\cite{Hinojosa2015}, and compute the winding numbers of the nodal points near the transition.
 In Sec. \ref{sec_4}  we consider a two-orbital/two band model of a \textit{d}-wave superconductor, which also undergoes a Lifshitz transition~\cite{Chubukov2016} when the pairs of nodal points along a symmetry direction merge and annihilate. We again compute winding numbers of these nodal points.
Finally, we make several concluding remarks in Sec. \ref{sec_6}.

\section{The Berry phase and the winding number}
\label{sec_2}

The topological properties of a system of interacting electrons in two dimensions (2D) are generally defined in terms of the Berry phase~\cite{Berry1984,Hasan2010, Bernevig2013}.
This phase reflects a non-trivial topological structure  of the wave function in the Hilbert space in the presence of topological defects, such as nodal points.
The Berry phase  $\gamma$,   is the phase which a wave function $\ket{n(\vec{R})}$ acquires when a system moves along a close path $\mathcal{C}$  around a topological defect in the space specified by
the set of parameters $\vec{R}$~ \cite{Fradkin2013, Bernevig2013}:
\begin{equation}
\gamma=-\oint_{\mathcal{C}} d\vec{R} \cdot A_{\vec{R}} = -\int_{\mathcal{S}} d\vec{S} \cdot B_{\vec{R}}, \nonumber
\end{equation}
Here $\mathcal{S}$ represents area in the parameter space, enclosed by the contour $\mathcal{C}$, $A_{\vec{R}}=-\mathrm{Im} \bra{n(\vec{R})} \nabla_{\vec{R}}\ket{n(\vec{R})}$,
and $B_{\vec{R}}=\nabla_{\vec{R}} \times A_{\vec{R}}$. The quantities $A_{\vec{R}}$ and $B_{\vec{R}}$ are called the Berry connection and
the Berry curvature. In our case, where the  parameter set is specified by momentum ${\bf k}$, the Berry phase is also called the Zak phase \cite{Zak1989}. The winding
number $Q$ is defined as the "normalized" Zak or Berry phase~\cite{Hasan2010,Fradkin2013,Bernevig2013}
\begin{equation}
Q=-\frac{1}{2 \pi} \oint_{\mathcal{C}} d\vec{k} \cdot A_{\vec{k}}=-\frac{1}{2 \pi} \int dk_x dk_y \left(\frac{\partial A_{k_y}}{\partial k_x}-\frac{\partial A_{k_x}}{\partial k_y} \right).
\label{Chern_n}
\end{equation}
In 2D systems this topological invariant represents an obstruction to the Stokes theorem and detects the presence of the nodal points~\cite{Bernevig2013,Sato2017}

 A standard recipe to obtain $Q$ for systems with  nodal points is to  expand the dispersion in the vicinity of
the node. A generic Hamiltonian near a nodal point can be cast into the form
\begin{equation}
H=-(k_i-k_i^0)A_{ij}\sigma_j,
\label{Dir_matr}
\end{equation}
where $k_i^0$ -- coordinates of the nodal point and $\sigma_j$ are the Pauli matrices. One can show~\cite{Bernevig2013} that the winding number
$ Q = - {\text sign} [\mathrm{det}(A)]$.

 In the next two sections we compute $Q$ for two models of FeSCs.
We first  compute $Q$ analytically and then verify the results numerically, using the computational method which has been proposed in Ref. \cite{Fukui2005}.

We will also explore a simple geometrical argument to compare the winding numbers for different nodal points.
Namely, suppose that there are two nodal points $1$ and $2$.  One can compute winding numbers $Q(1)$ and $Q(2)$ by integrating along two different contours, each surrounding only one nodal point. Both contours should have the \textit{same} direction of bypass. Alternatively, one can
transform the coordinates, separately  for region $1$ and region $2$, and bring the nodes to the same point in space.
The integration contours then become the same, modulo the direction of the bypass. The  winding numbers $Q(1)$ and $Q(2)$ then are be the same if the bypass directions in new basis is be the same, or  have opposite sign if the  bypass directions in new basis are opposite.

\section{An  $\mathit{s}$-wave superconductor with accidental nodes}
\label{sec_3}

\subsection{The model}
\label{subsec_3}

We consider a 2D model of an FeSC with  hole pockets centered at $\Gamma = (0,0)$ and  electron pockets centered at $(0,\pi)$ and $(\pi, 0)$ in 1Fe Brillouin zone (BZ).
We assume  that the dominant interaction is in the $s$-wave ($A_{1g}$) channel, and the system develops an $s^{+-}$ superconductivity  with $\pi$ phase difference between the gaps on hole and electron pockets.  The gaps on $\Gamma$-centered hole pockets are $C_4$ symmetric, with $\cos{4n \theta_h}$ variation  along the hole pockets. The electron pockets are centered at non-$C_4$-symmetric points, and the gap variation along the electron pockets has additional $\pm \cos{(4m+2) \cos{\theta_e}}$ components (with $\theta_e$ counted from the same axis on both electron pockets).  We assume, following earlier works, that the $\cos{2\theta_e}$ variation  is the strongest one, and it gives rise to accidental nodes on the electron pockets.  The gap on hole pockets has no nodes, and we will not include hole pockets into our consideration.

The position of the accidental nodes can be manipulated by including the hybridization between the two electron pockets~\cite{Hinojosa2015,Khodas2012}.  The hybridization is caused by  pnictogen/chalcogen atoms, which are located  above and below an Fe plane, in "up-down" order.  As a result, the actual unit cell is bigger and contains 2 Fe atoms. One can still work in a 1Fe unit cell, but there the hybridization gives rise to terms in the Hamiltonian, in which incoming and outgoing  momenta differ by $(\pi,\pi)$.   In a superconductor, there are two types of such  terms -- one describes the hopping  between the electron pockets, another describes a creation or annihilation of Cooper pairs made of fermions from different electron pockets. Both terms affect the position of the gap nodes.  For definiteness, here we consider the effects due to additional pairing terms induced by the  hybridization.

The Hamiltonian of the model is
\begin{equation}
H=H_0+H_{\Delta}+H_{\beta}
\label{initH}
\end{equation}
where
\begin{equation}
H_0=\sum_k \xi^c_k c^{\dag}_{k\alpha} c_{k\alpha} + \xi^d_k d^{\dag}_{k\alpha} d_{k\alpha}
\end{equation}
is the kinetic energy of fermions near the two electron pockets,
\begin{equation}
H_{\Delta}= \frac{1}{2} \sum_k \left[ \Delta_c ~c^{\dag}_{k\alpha}c^{\dag}_{-k\beta}+ \Delta_d ~ d^{\dag}_{k+Q\alpha}d^{\dag}_{-k-Q\beta} \right]i\sigma^y_{\alpha\beta},
\end{equation}
is the pairing term with angle-dependent gap functions $\Delta_c = \Delta (1-y_e), ~\Delta_d = \Delta (1 +y_e)$, where  $y_e=\alpha \, \mathrm{cos}2\theta_e$, and $\alpha$ is a parameter, which depends on the orbital composition of electron pockets.  When $\alpha >1$,  $\Delta_c$ and $\Delta_d$ have accidental nodes.
 Finally,
\begin{equation}
H_{\beta}=\frac{1}{2} \sum_k \beta \left[ c^{\dag}_{k\alpha}d^{\dag}_{-k-Q\beta} + d^{\dag}_{k+Q\alpha}c^{\dag}_{-k\beta} \right]
i\sigma^y_{\alpha\beta}
\end{equation}
 is the additional pairing term, induced by the hybridization, in which the total momentum of the pair equals to $(\pi,\pi)$.  Without the loss of generality we set $\beta$ to be positive.  We will see that by varying the strength of $\beta$ one can move the positions of the accidental nodes.

 It is instructive to consider separately the special case, when the electron pockets can be approximated as circular, and a generic case, when they are elliptical.
 For both cases we assume that $\alpha >1$, i.e., in the absence of hybridization the gap functions $\Delta_c$ and $\Delta_d$ have accidental nodes.

\subsubsection{Circular pockets}

\begin{figure}[h]
\center{\includegraphics[width=1\linewidth]{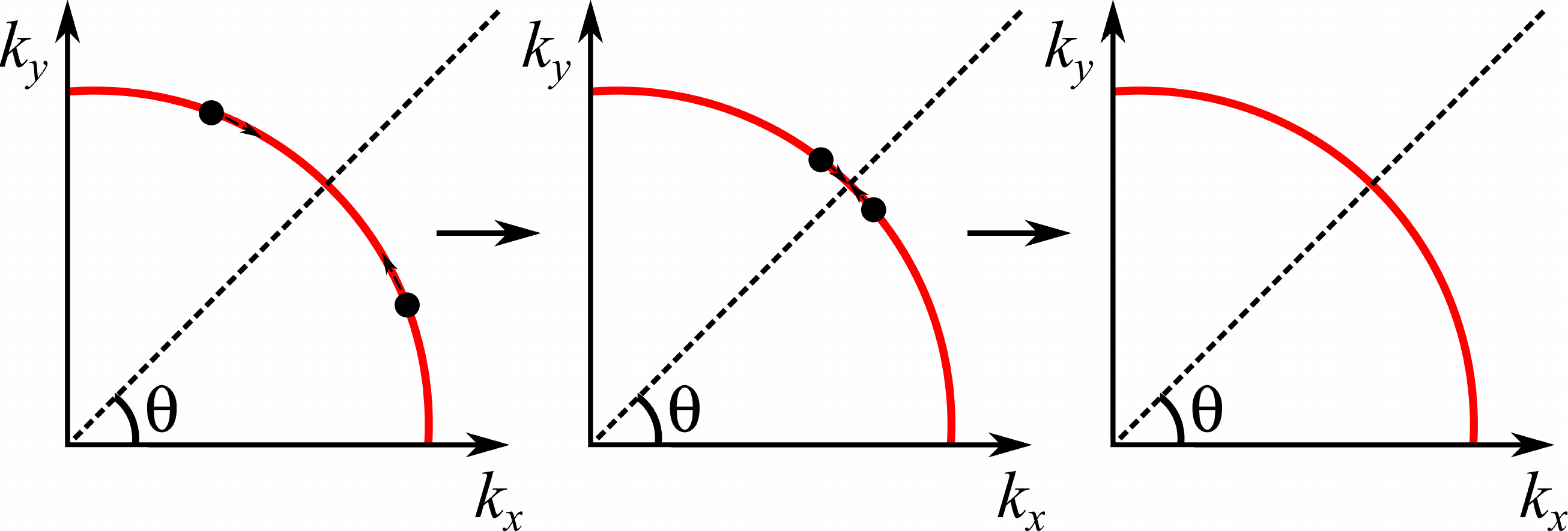}}
\centering{}\caption{The location of the nodal points on one of the FS of hybridized circular electron pockets.  Red line -- the first quadrant of the  FS,
  black dots -- the  nodal points.  The nodes move towards BZ diagonals (in 1Fe zone) with increasing the  strength of the hybridization parameter $\beta$.
  The neighboring nodal points merge and disappear at $\beta_{\mathrm{crit}}=\Delta$. }
\label{sk}
\end{figure}

For circular electron pockets
$\xi_k^c=\xi_k^d=\xi_k$. The Hamiltonian (\ref{initH}) can be straightforwardly diagonalized by Bogoliubov transformation to~\cite{Hinojosa2015}
\begin{equation}
H= E_0 + \sum_{k,\alpha}  E_k^{+} e^{\dag}_{k\alpha} e_{k\alpha} + E_k^{-} f^{\dag}_{k\alpha} f_{k\alpha}
\end{equation}
where
\begin{equation}
\left(E_k^{\pm}\right)  = \left[\xi_k^2 + \left( \Delta \pm \sqrt{\Delta^2 y_k^2 + \beta^2}  \right)^2 \right]^{1/2}
\label{cd}
\end{equation}
The dispersion $E^+$ is obviously nodeless,  but $E^-$  has zeros at
\begin{equation}
\mathrm{cos} (2 \theta_e) = \pm \frac{\sqrt{\Delta^2 - \beta^2}}{\alpha \Delta}.
\end{equation}
At $\beta<\beta_{\mathrm{crit}}=\Delta$ there are 8 nodal points, two in each of the four quadrants.   At the critical value $\beta=\beta_{\mathrm{crit}}$ the pairs of nodal points merge along the BZ diagonals. At $\beta>\beta_{\mathrm{crit}}$, the nodes disappear (see Fig. \ref{sk}).

\subsubsection{Elliptical pockets}

\begin{figure}[h]
\center{\includegraphics[width=1\linewidth]{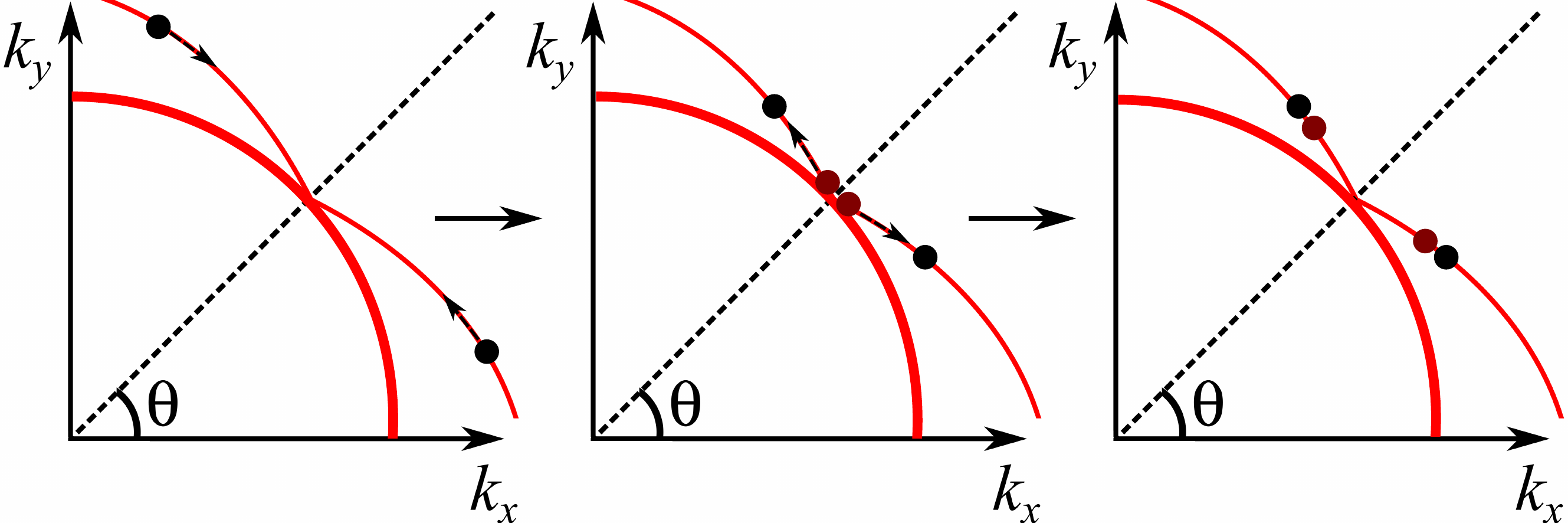}}
\centering{}\caption{The location of the nodal points on one of the FS of hybridized  elliptical electron pockets, for large enough degree of ellipticity. Thick red line -- the first quadrant of the  FS, thin red line -- the locus of location of the nodal points, black dots -- the original nodal points.   As the hybridization parameter $\beta$ increases, nodal points move towards the BZ  diagonal, but don't reach it. Instead, at  $\beta=\Delta$, a new pair of nodal points (brown dots)  appears along BZ diagonal, and at larger  $\beta$
    move towards the existing nodes. The old and the new nodal points merge and disappear at $\beta=\beta_{\mathrm{crit}} > \Delta$.}
\label{sk2}
\end{figure}

For elliptical pockets the dispersions are
\begin{equation}
\xi^c_k = \frac{k_x^2}{2m_1} + \frac{k_y^2}{2m_2}-\mu, \; \;  \xi^d_k = \frac{k_x^2}{2m_2} + \frac{k_y^2}{2m_1}-\mu,
\end{equation}
expanding near the Fermi surface we obtain~ \cite{Vorontsov2010,Hinojosa2015}
\begin{equation}
\begin{split}
\xi_k^{c,d}=\xi_k \pm \delta \, \mathrm{cos} (2\theta_k), \\
\delta \approx k^2_F  \frac{m_2-m_1}{4 m_1 m_2},~  \xi_k = k^2/2m^* - \mu, ~ m^* =2m_1 m_2/(m_1+m_2)
\end{split}
\label{disp}
\end{equation}
 Diagonalizing the Hamiltonian we again obtain two bands  with the dispersion
 $(E^{\pm}) =\left(A_k\pm \sqrt{B_k} \right)^{1/2}$, where
\begin{equation}
\begin{gathered}
A_k=\frac{1}{2} \left[(\xi^c_k)^2 + (\xi^d_k)^2 + 2\Delta^2(1+y_k^2) + 2\beta^2 \right], \\
B_k=\frac{1}{4} \left[\left( (\xi^d_k)^2 - (\xi^c_k)^2 + 4 \Delta^2 y_k \right)^2 + 4 |\beta|^2 \left( (\xi^c_k -\xi^d_k)^2 + 4\Delta^2 \right) \right].
\end{gathered}
\end{equation}
Using Eqs. (\ref{disp}) we can rewrite $(E^{-})$ as
\begin{equation}
\begin{split}
\left(E^{-}\right) =\biggr[ \xi^2_k + \Delta^2 + \beta^2 + \mathrm{cos}^2  (2\theta) \left( \delta^2 + \Delta^2 \alpha^2 \right) - \\
2\sqrt{\mathrm{cos}^2 (2\theta) \left(\Delta^2 \alpha - \xi \delta_k  \right)^2 + |\beta|^2 \left(\delta^2  \mathrm{cos}^2 (2\theta) + \Delta^2 \right)} \biggr]^{1/2}.
\end{split}
\label{spectr}
\end{equation}
In distinction to circular pockets, nodal points are now located not on the original Fermi surface, but at
\begin{equation}
\begin{gathered}
\xi=\frac{\delta^2 - \alpha^2 \Delta^2 \pm \sqrt{\left( \alpha^2 \Delta^2 + \delta^2 \right)^2 - 4 \alpha^2 \beta^2 \delta^2}}{2|\alpha| \delta}, \\
\mathrm{cos}^2 (2\theta)=\frac{\delta^2 - \alpha^2 \Delta^2 \mp \sqrt{\left( \alpha^2 \Delta^2 + \delta^2 \right)^2 - 4 \alpha^2 \beta^2 \delta^2}}{2\alpha^2 \delta^2}. \\
\end{gathered}
\label{sol}
\end{equation}
A straightforward analysis shows~\cite{Hinojosa2015} that the evolution of the nodal points with increasing $\beta$ depends on the interplay between the ellipticity  parameter $\delta$ and
$\alpha \Delta$. When $\delta < \alpha \Delta$, pairs of nodes in each quadrant  merge and disappear at $\beta=\Delta$ on
the diagonals of the BZ, like in the case of circular pockets.  When $\delta> \alpha \Delta$,  nodal points don't reach diagonals when $\beta$ reaches $\Delta$.  At this $\beta$, a new pair on nodes appears along each diagonal (see Fig. \ref{sk2}). As $\beta$ continues increasing, the new nodal points move towards the existing nodes.
The new and old nodes merge and disappear at the critical
\begin{equation}
\beta_{\mathrm{crit}}= \frac{\alpha^2 \Delta^2 + \delta^2}{2 |\alpha | \delta} > \Delta.
\end{equation}

\subsection{The winding number}

\subsubsection{Circular pockets}

\begin{figure}[h!]
\begin{minipage}[h]{0.95\linewidth}
\center{\includegraphics[width=0.95\linewidth]{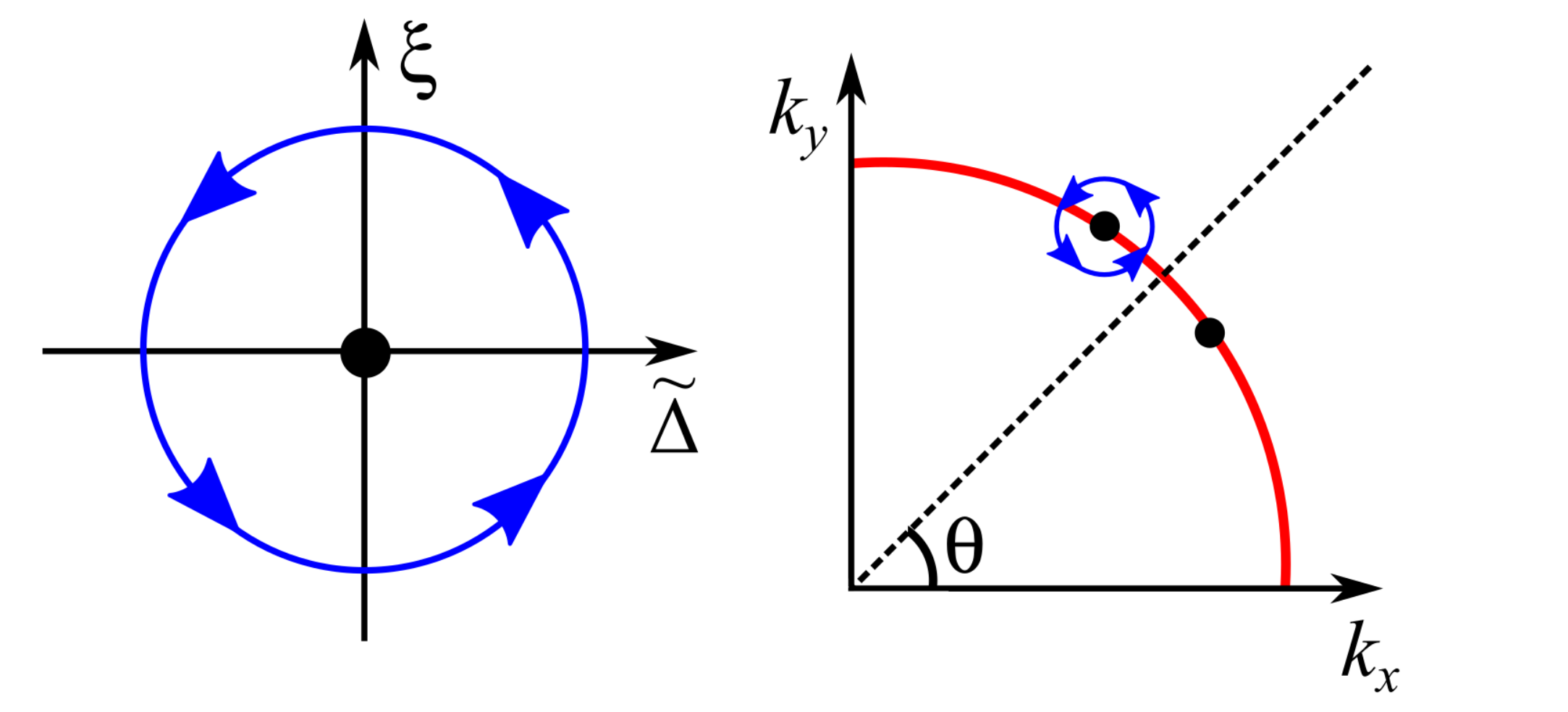} }
\end{minipage}
\vfill
\begin{minipage}[h]{0.95\linewidth}
\center{\includegraphics[width=0.95\linewidth]{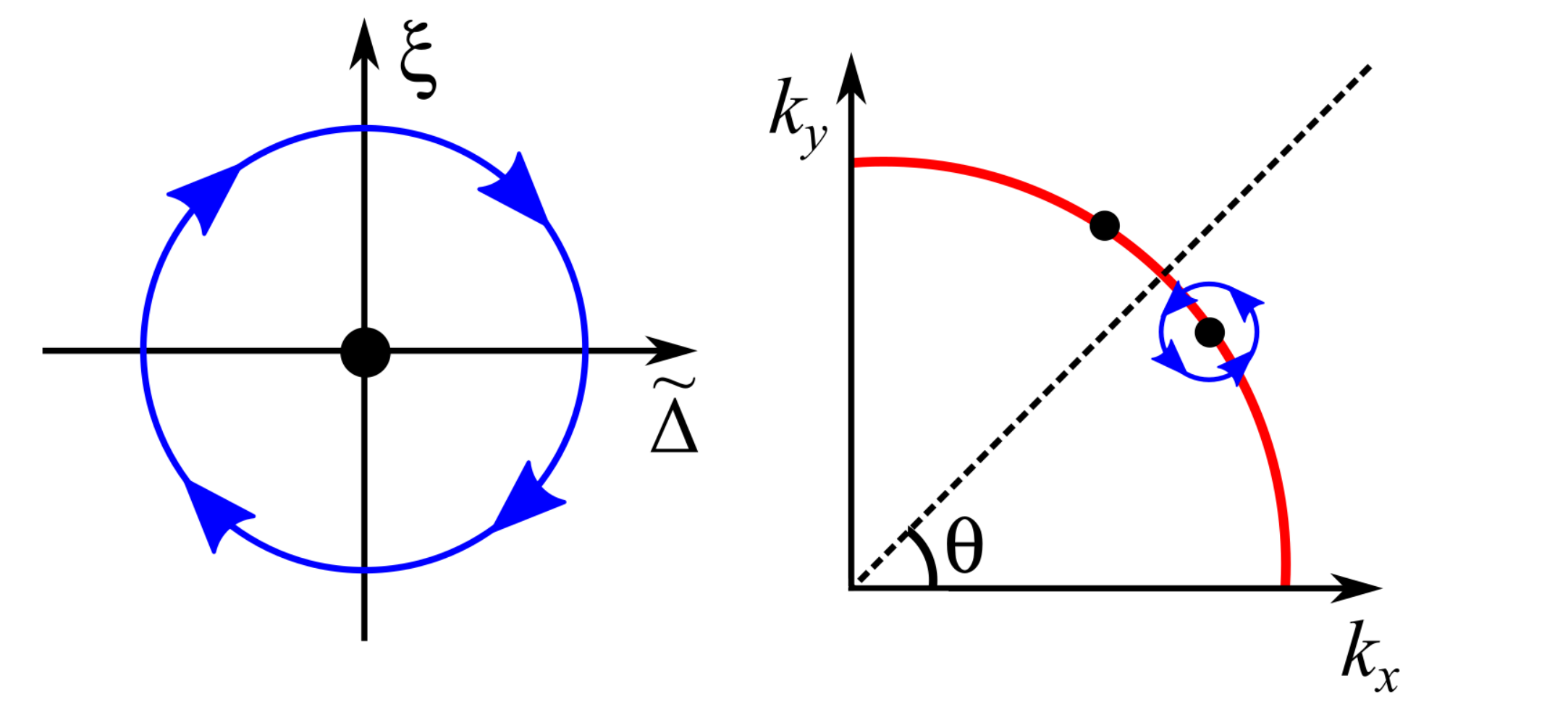} }
\end{minipage}
\centering{}\caption{Winding around each of the two nodal points shown in the right panel.  Both points are mapped to  the origin of the coordinates by transforming to the new basis with variables $(\xi, \tilde{\Delta})$ instead of $(k_x, k_y)$ (left panel).  Different winding numbers for the two nodes on the right panel and
due to different directions of bypass on the left panel.}
\label{contours_circ}
\end{figure}

To obtain the winding numbers for the nodal points we expand $E^-_k$  in Eq. (\ref{cd}) near each of 8 nodal points. Because of $C_4$ symmetry, we only consider the first quadrant $\theta \in [0, \pi/2]$. The dispersion (\ref{cd}) near a nodal point has the Dirac form
\begin{equation}
E_k^{\mathrm{Dirac}} = \sqrt{\left( \frac{d \xi}{dk}(k-k') \right)^2 + \left( \frac{2 \Delta^2 \alpha^2 \mathrm{sin}(2\theta) \mathrm{cos}(2\theta)}{\sqrt{\Delta^2 \alpha^2 \mathrm{cos}^2 (2\theta) + \beta^2}} (\theta-\theta') \right)^2},
\end{equation}
where $k' = k_F$ and $\theta'$ are the coordinates of $E^-_k =0$.
The corresponding Dirac Hamiltonian can be obtained using the Pauli matrices
\begin{equation}
H^{\mathrm{Dirac}} =\frac{d\xi}{dk}(k-k') \cdot \sigma_3 +  \frac{\Delta^2 \alpha^2 \mathrm{sin}(4\theta)}{\sqrt{\Delta^2 \alpha^2 \mathrm{cos}^2 (2\theta) + \beta^2}} (\theta-\theta') \cdot \sigma_1,
\end{equation}
or, in the explicit matrix form,
\begin{equation}
H^{\mathrm{Dirac}} = \begin{pmatrix}
\frac{d\xi}{dk}(k-k') & \frac{\Delta^2 \alpha^2 \mathrm{sin}(4\theta) }{\sqrt{\Delta^2 \alpha^2 \mathrm{cos}^2 (2\theta) + \beta^2}} (\theta-\theta')\\
\frac{\Delta^2 \alpha^2 \mathrm{sin}(4\theta) }{\sqrt{\Delta^2 \alpha^2 \mathrm{cos}^2 (2\theta) + \beta^2}} (\theta-\theta') & -\frac{d\xi}{dk}(k-k')
\end{pmatrix}.
\label{Dirac_circ}
\end{equation}
We associate $\theta$ with the first nodal direction and $k$ with the third one, and rewrite Dirac Hamiltonian $H^{\mathrm{Dirac}}$ in the form of
Eq. (\ref{Dir_matr}) with $A_{ij}$ ($i,j =1,3$)
\begin{equation}
A = \begin{pmatrix}
\frac{\Delta^2 \alpha^2 \mathrm{sin}(4\theta)}{\sqrt{\Delta^2 \alpha^2 \mathrm{cos}^2 (2\theta) + \beta^2}} &0 \\
 0 & \frac{d\xi}{dk}
\end{pmatrix}.
\end{equation}
The sign of the det $A$ depends only on sign of $\mathrm{sin}(4\theta)$, which is positive at $\theta<\pi/4$ and negative at $\theta>\pi/4$. Nodal points are located  on the opposite sides of $\theta=\pi/4$, hence their winding numbers are opposite: -1 and +1.

We can obtain the same result by introducing the effective pairing Hamiltonian for fermions with $E^{-}_k$ in the form
\begin{equation}
H_{eff} = \begin{pmatrix}
\xi & \tilde{\Delta}  \\
\tilde{\Delta} & -\xi
\end{pmatrix}
\end{equation}
with $\tilde{\Delta}=\Delta - \sqrt{\Delta^2 y_k^2 + \beta^2}$, and treating  $(\xi,\tilde{\Delta})$ as new effective coordinates.  The transformation from $(k_x,k_y)$ to $(\xi,\tilde{\Delta})$ is multi-valued: all 8 solutions for $E^-_k=0$ are
now mapped to the origin in $(\xi,\tilde{\Delta})$-plane. Then the contour $C$ in Eq. (1) is the same for all nodal points,
and the signs of the winding numbers depend only on the bypass direction of $C$ for a given node (which is a topological invariant).  Because
 $\tilde{\Delta}$ depends on $\cos {2\theta}$, we have different direction of bypass for each pair of nodal points in a given  quadrant
  (see Fig. \ref{contours_circ}). Indeed, consider the nodal point located between $\theta=0$ and $\pi/2$. Let's choose the counterclockwise bypass along the closed contour in the $(k_x,k_y)$-plane.  The bypass  starts at $\xi=0, \, \theta>\theta_{\mathrm{sol}}$ and goes to the point $\xi<0, \, \theta=\theta_{\mathrm{sol}}$, where $\theta_{\mathrm{sol}}$ is the solution for $\tilde{\Delta}=0$. One can easily verify that this corresponds to clockwise bypass direction in the $(\xi,\tilde{\Delta})$-plane. Using the same strategy, one can then verify that the same  bypass in the $(k_x,k_y)$-plane for another nodal point (the one with larger $\theta_{\mathrm{sol}}$) corresponds to counterclockwise direction in the  $(\xi,\tilde{\Delta})$-plane. This obviously  gives the opposite sign of the winding number.

\subsubsection{Elliptical pockets}

We now extend the analysis to elliptical pockets.
We expand the Hamiltonian of Eq. (\ref{initH}) in Taylor series in the vicinity of each nodal point and obtain  the Dirac Hamiltonian in the form
\begin{equation}
H = \begin{pmatrix}
\frac{dE^-}{dk^2} (k^2-(k')^2) & \frac{dE^-}{d\theta}(\theta-\theta')\\
\frac{dE^-}{d\theta}(\theta-\theta') & -\frac{dE^-}{dk^2} (k^2-(k')^2)
\end{pmatrix}.
\end{equation}
Associating $\theta - \theta'$ and $k^2-(k')^2$ with the directions $i=1$ and $i=3$, respectively, we obtain the matrix $A$ in Eq. (\ref{Dir_matr}), as
\begin{equation}
A = \begin{pmatrix}
 \frac{dE^-}{d\theta}& 0\\
0& \frac{dE^-}{dk^2}
\end{pmatrix},
\label{A_ell}
\end{equation}
where
\begin{equation}
\begin{gathered}
\frac{dE^-}{dk^2} = \frac{X-Y \mathrm{cos}^2(2\theta)}{\sqrt{D}} \\
\frac{dE^-}{d\theta} = \frac{Z \mathrm{sin}(4\theta)}{\sqrt{D}}
\end{gathered}
\end{equation}
and
\begin{equation}
\begin{gathered}
D = \beta^2 + \Delta^2 + \xi_k^2 + \left(\delta^2 + \alpha^2 \Delta^2 \right) \mathrm{cos}^2(2\theta) \\ - 2\sqrt{\beta^2\Delta^2 + \left( \beta^2 \delta^2 + (\alpha \Delta^2 - \delta \xi_k)^2 \right)\mathrm{cos}^2(2\theta)}, \\
X = 2c_1\xi_k+2 c_2 \delta \mathrm{cos}^2(2\theta), \\
Y = \frac{2 c_2  \left( 2c_1^2 k^4 \delta+\alpha\Delta^2\mu - c_1k^2(2\alpha\Delta^2 + 3\delta \mu) +\delta (\beta^2 + \mu^2) \right)}{\sqrt{\beta^2 \Delta^2 + \left(\beta^2 \delta^2 + (\alpha \Delta^2 - \delta \xi_k)^2 \right) \mathrm{cos}^2(2\theta)}}, \\
Z = \left(-\delta^2 - \alpha^2 \Delta^2 + \frac{\beta^2 \delta^2 + (\alpha \Delta^2 - \beta \xi_k)^2}{\sqrt{\beta^2 \Delta^2 + \left(\beta^2 \delta^2 + (\alpha \Delta^2 - \delta \xi_k)^2 \right) \mathrm{cos}^2(2\theta)}} \right).
\end{gathered}
\end{equation}
Here we introduced
 $c_1=\frac{1}{4}\left(\frac{1}{m_1}+\frac{1}{m_2}\right)$ and $c_2=\frac{1}{4}\left(\frac{1}{m_1}-\frac{1}{m_2}\right)$.

For brevity, we focus on the case when ellipticity is strong enough ($\delta > \alpha \Delta$, see Sec. \ref{subsec_3}) and consider the winding numbers for the two nodal points, which emerge at $\beta = \Delta$ along the diagonals,  and then move merge with the existing nodal points at $\beta= \beta_{crit} > \Delta$.

Because of $C_4$ rotational symmetry, we again focus on the nodal points at $0 < \theta < \pi/4$.
We computed  the determinant of (\ref{A_ell}) numerically for $\Delta=1, \beta=1.002, \mu=10, \alpha=-1.5, m_2/m_1=2,$ and verified that the winding numbers for the  new nodal point, which appears at  $\beta = \Delta$, and the "old" nodal point, with which the new one eventually merges,  have opposite signs of the winding number.

We next discuss the computation of the winding numbers in the "geometrical" approach, when we transform different nodal points into the same location.  For $\beta \geq \Delta$, the two emerging nodal points are still close to the diagonals, and  we can expand $E^{-}_k $ in powers of $\xi$ and $\cos{2 \theta}$. The expansion yields
\begin{equation}
(E_k^{-})^2 \approx E^2_{\mathrm{lin}}=\xi^2+  F(\theta),
\label{e1}
\end{equation}
where
$\xi$ is a function of $\beta$ from Eq. (\ref{sol}) (one should choose the solution for which $\xi =0$ for $\beta = \Delta$), and
\begin{equation}
\begin{split}
F(\theta)=\mathrm{cos}^2(2\theta) \left[\delta^2 + \Delta^2 \alpha^2 - \frac{(\Delta^2 \alpha - \xi \delta)^2 + |\beta|^2 \delta^2}{|\beta| \Delta}  \right] \\
+\mathrm{cos}^4(2\theta)  \frac{\left[(\Delta^2 \alpha - \xi \delta)^2 + |\beta|^2 \delta^2\right]^2}{4|\beta|^3 \Delta^3}+ (\Delta-|\beta|)^2 .
\end{split}
\end{equation}

\begin{figure}[h]
\center{\includegraphics[width=0.9\linewidth]{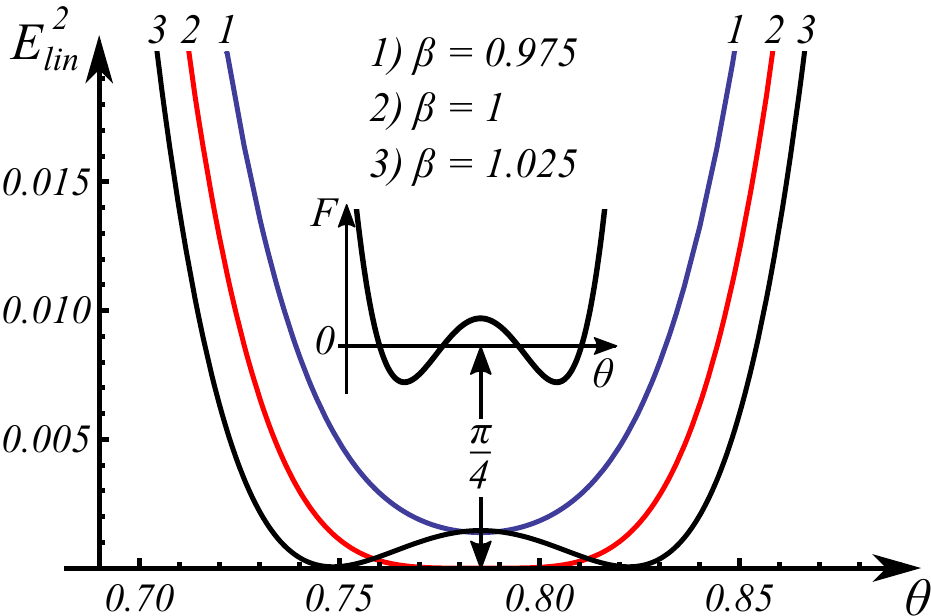}}
\centering{}\caption{Angular dependence of $E^2_{\mathrm{lin}}$, Eq. (\ref{e1}),  for three values of $\beta$, smaller, equal, and larger than $\Delta$, which we set equal to one in proper units.  Inset: $F(\theta)$ as a function of the angle $\theta$ for $\beta=1.025 > \Delta$. The minima of $F$ correspond to the locations of emerging nodal points. We set $\mu=10, \alpha=-1.5, m_2/m_1=2.$}
\label{enerlin}
\end{figure}

We plot $E_{\mathrm{lin}}^2$ as a function of $\theta$ in Fig. \ref{enerlin}. The new nodal points emerge at  $\beta =\Delta$, at $\xi = F=0$.
As $\beta$ increases, the two nodal points  split and move  towards already existing nodal points.

\begin{figure}[h]
\begin{minipage}[h]{0.95\linewidth}
\center{\includegraphics[width=0.95\linewidth]{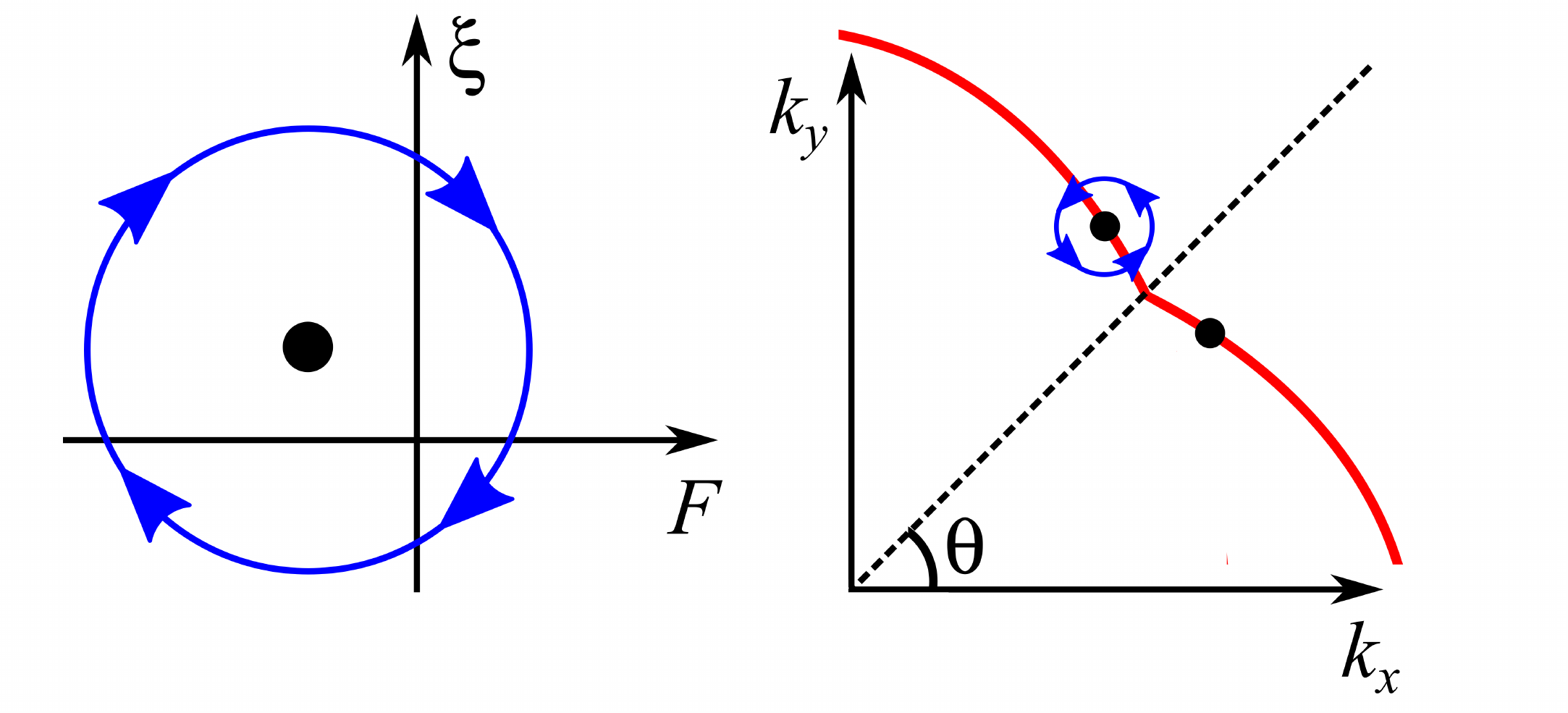} }
\end{minipage}
\vfill
\begin{minipage}[h]{0.95\linewidth}
\center{\includegraphics[width=0.95\linewidth]{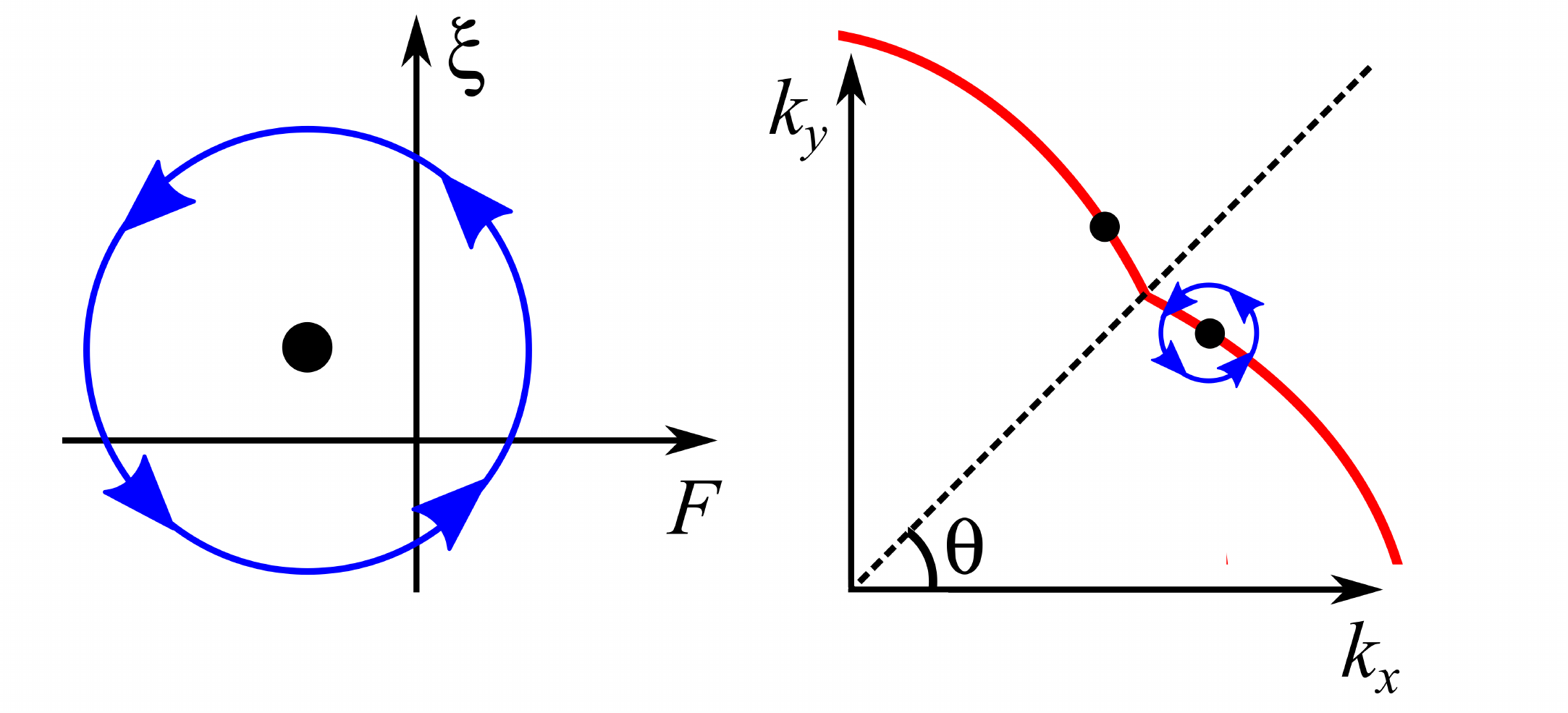} }
\end{minipage}
\centering{}\caption{The case of elliptical pockets.  Bypass trajectories around the emerging nodal points (black dots)  at $\beta > \Delta$ for two sets of coordinates: $(k_x, k_y)$ in the right panel and  $(\xi, F)$ in the left panel (see the text for the definitions of $\xi$ and $F$). In left panel, the two nodal points are mapped into the same point in the new coordinates. Opposite signs of the winding numbers for these two points are due to different directions of bypass, as shown in the left panel.}
\label{contours}
\end{figure}

To calculate the winding numbers of these nodal points we transform to the $(\xi,F)$ plane, where the two nodal points are moved to the same $\xi$ and $F$.
 In distinction to the case of circular pockets, the  nodal points are now located at finite $\xi$ and $F$,  given by  $E_{\mathrm{lin}}=0$.
  Still, the integration contour $C$ is the same for both nodal points,  and one can extract the  winding numbers from the bypass directions.
Consider the nodal point in the upper panel of Fig. \ref{contours}.  Let us choose the counterclockwise bypass along the closed contour in the $(k_x,k_y)$-plane. In $(\xi, F)$ plane, this  bypass starts at $\xi=0, \, F(\theta)>0$, proceeds  to the point $\xi<0, \, F(\theta)=0$ and then reaches $\xi=0, \, F(\theta)<0$. This is clockwise bypass in the $(\xi,F)$-plane. For the nodal point in the lower panel of Fig. \ref{contours}, the same consideration shows that  the bypass direction in the $(\xi,F)$-plane changes to counterclockwise. This implies that the winding numbers for the two emerging nodal points are opposite.

The winding numbers of the original nodal points can be obtained in the same way as was done for circular pockets  because these nodal points survive when the ellipticity parameter $\delta$ vanishes. Comparing the  directions of bypass  in Figs. \ref{contours_circ} and \ref{contours} we see that the nodal points, which eventually merge and disappear, always have the winding numbers of opposite sign.

\subsection{Numerical analysis}

\subsubsection{The numerical procedure}

We supplement our analytical calculations with the numerical analysis. We use the computational procedure  introduced in Ref. \cite{Fukui2005}.
It uses discrete grid functions for Berry connection and Berry curvature. In order to calculate these functions,  one has to define the wave function of the system.  In  Nambu notation,  a  field operator is $\Psi=(v^c_k c_k^{\dagger} + u^c_k c_{-k} + v^d_k d_k^{\dagger} + u^d_k d_{-k})$, where $u^c_k, v^c_k, u^d_k, v^d_k$ are  Bogoliubov transformation coefficients. The wave function of the system $\ket{n({\bf k})}$ is then a  4-component vector \cite{Heinzner2005} made out of  Bogoliubov coefficients. We need the two wave functions which correspond to eigenvalues $ \pm E^- (k)$, which, we remind, describe the excitation branch with the nodes.

For the numerical computation of the Berry curvature, we follow Ref. \cite{Fukui2005} and introduce the grid on the BZ, i.e., coarse-grain momenta to
 $\vec{k}=k_{ij} = \left(2 \pi i / a_x, \, 2 \pi j /a_y\right)$, where $a_x, a_y$ are grid spacings, each a fraction of the interatomic spacing.   It was argued that the value of $B$ doesn't depend on grid spacing as long as each elementary cell contains no more than one nodal point. We next introduce a link variable  $\tilde{A}_{\vec \delta} (\vec{k})$  on a grid (a "discrete Berry connection"):
\begin{equation}
\tilde{A}_{\vec{\delta}} (\vec{k})=\langle n(\vec{k}) | n(\vec{k}+\vec{\delta}) \rangle/N.
\end{equation}
where $N = |\langle n(\vec{k}) | n(\vec{k}+\vec{\delta}) \rangle|$ -- is the normalization factor, and $\delta_x =  (2 \pi / a_x, \, 0)$, $\delta_y = (0, \, 2 \pi /a_y)$. This $\tilde{A}_{\vec{\delta}} (\vec{k})$ determines the phase, which $\ket{n(\vec{k})}$ acquires under the change from $\vec{k}$ to $\vec{k} +\vec{\delta}$.
The total phase change over an elementary closed loop adjacent to a particular $\vec{k} = k_{ij}$ (i.e., a particular combination of $i, j$) is
\begin{equation}
K(\vec{k}) =  \tilde{A}(\vec{k})_{\vec{\delta}_x} \tilde{A}(\vec{k}+\vec{\delta}_x )_{\vec{\delta}_y} \tilde{A}(\vec{k}+\vec{\delta}_y )_{\vec{\delta}_x}^{-1} \tilde{A} (\vec{k})_{\vec{\delta}_y}^{-1}.
\end{equation}
Taking the logarithm of $K$ we obtain the phase change over a loop:
\begin{equation}
\tilde{B}(\vec{k}) = \frac{1}{i} \mathrm{ln} K(\vec{k}) = \phi(\vec{k}).
\label{log}
\end{equation}
If there is no node inside a loop for a given $\vec{k} = \vec{k}_0$, the overall phase change is zero. If a given loop encircles a nodal point, then, within the loop, one moves from the lower to the upper branch of the Dirac spectrum (or vise versa), and the phase changes by $\pm 2\pi$. Accordingly, $\tilde{B}(\vec{k}_0)/2\pi$ gives the winding number of this nodal point. In the ideal situation, $\tilde{B}(\vec{k})$ will be non-zero only for a discrete set of $\vec{k}_0$, equal to the number of nodal points.  In numerical calculations, however,  the logarithm in Eq. (\ref{log})  often strongly oscillates between $2\pi$ and $-2\pi$,  if a nodal point is near the trajectory along the loop. To avoid this complication, we add to the Hamiltonian the term $m \sigma^y$ and compute $\tilde{B}(\vec{k})$ for all $\vec{k}$ in the BZ.  This term makes the value of the logarithm  well defined, but at the same time, it couples lower and upper branches of the Dirac spectrum, and, as a result, $\tilde{B}(\vec{k})$ becomes non-zero for all ${\bf k}$ in the BZ.
    Still, as long as $m$ is small, the numerics clearly shows an enhancement of the magnitude of $\tilde{B}$ near a node. Because our primary goal is to check the signs of the winding numbers, it is sufficient to compute $\tilde{B}(\vec{k})$ for a small but finite $m$ and check the sign of $\tilde{B}(\vec{k})$  near each nodal point.

\subsubsection{Circular pockets}
\begin{figure}[h]
\center{\includegraphics[width=0.8\linewidth]{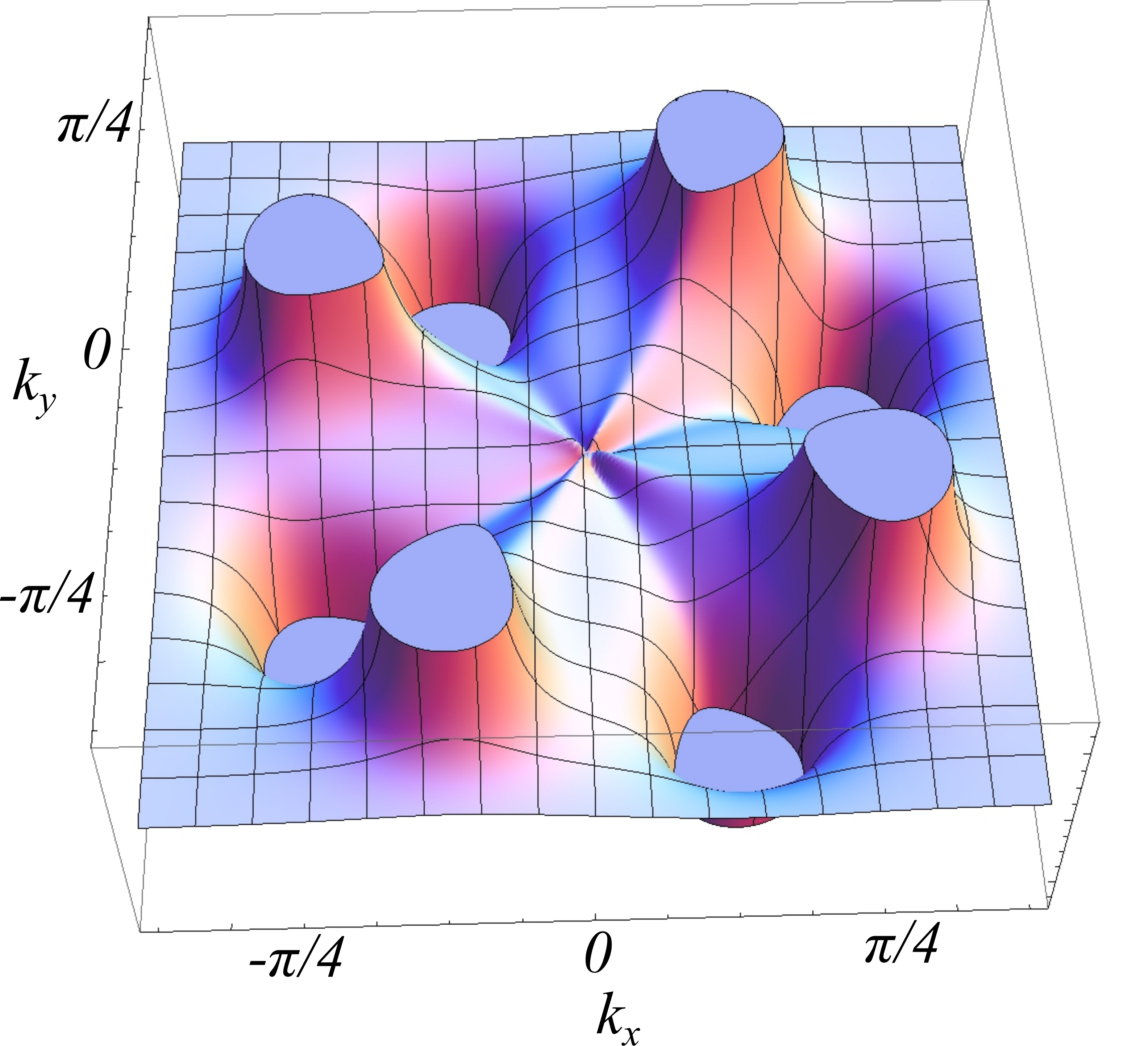}}
\centering{}\caption{The grid Berry curvature $\tilde{B}$ for the case of circular electron pockets. We used $\beta=0.5, \mu=1, \Delta=1, \alpha=1.5$.
The grid shows the checkerboard order of positive and negative winding numbers for the eight nodal points.}
\label{circ}
\end{figure}

 The results  of our calculations of $\tilde{B}(\vec{k})$ for circular pockets are shown in Fig. \ref{circ}.
We found  that eight nodal points have the winding
numbers $\pm1$. This is fully  consistent with the analytical result.
We also see from Fig. \ref{circ} that there is a checkerboard order of nodal points with positive and negative values of the  winding number.  This is again consistent with the analytical results.

\subsubsection{Elliptical pockets}

\begin{figure}[t]
\center{\includegraphics[width=0.9\linewidth]{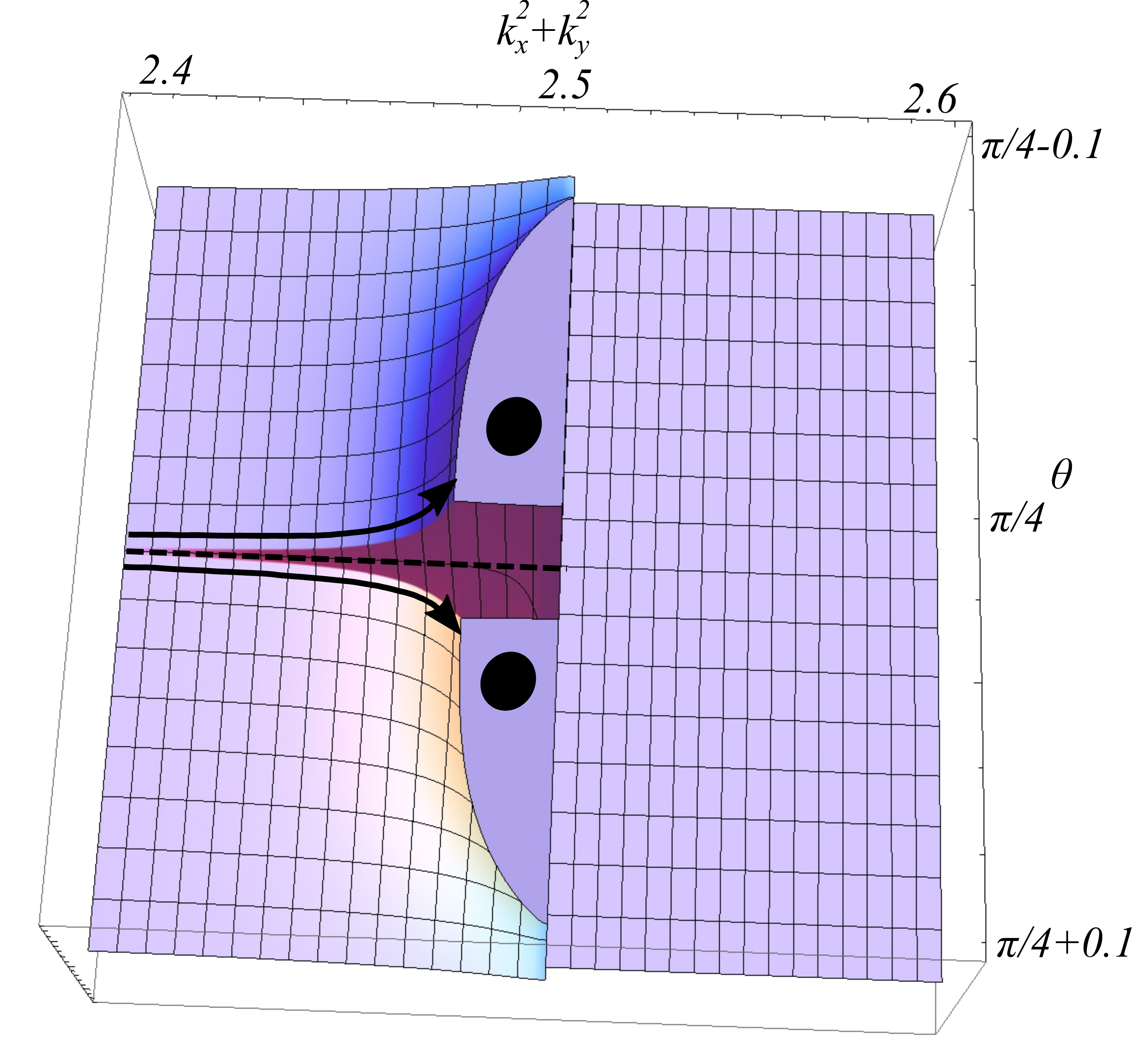} }
\centering{}\caption{The grid Berry curvature $\tilde{B}$ for the two emerging  nodal points in the case when the electron pockets are elliptical. The positions of the nodal points are  shown by black dots. Unusual shape of the plateaus around nodal points is caused by the choice of polar coordinates. The arrows show where the grid Berry curvature is positive, and where it is negative.  We used $\mu=10, \alpha=-1.5, \beta=1.01, \Delta=1, m_2/m_1=2.$}
\label{ellipt}
\end{figure}

For elliptical pockets,  we used as the point of departure the effective band Hamiltonian  representing the low-energy band $E^{-}$, which has nodal points. We introduce a $2 \times 2$ matrix  Hamiltonian,  which  gives the dispersion in Eq.  ($\ref{spectr}$)
and apply the numerical procedure, described above. We plot the Berry curvature as a function of $|\vec{k}|$ and $\theta$ in Fig. \ref{ellipt}.
 As we can see from this figure, in the region around the nodal points, the  Berry curvature saturates at a positive value near one point and at  negative value near the other.
 This leads to opposite signs of the winding numbers around these points.  This is again fully consistent with the analytical results.

\section{A two-band $\mathit{d}$-wave superconductor}
\label{sec_4}

\subsection{The model}

\begin{figure}[h]
\center{\includegraphics[width=1\linewidth]{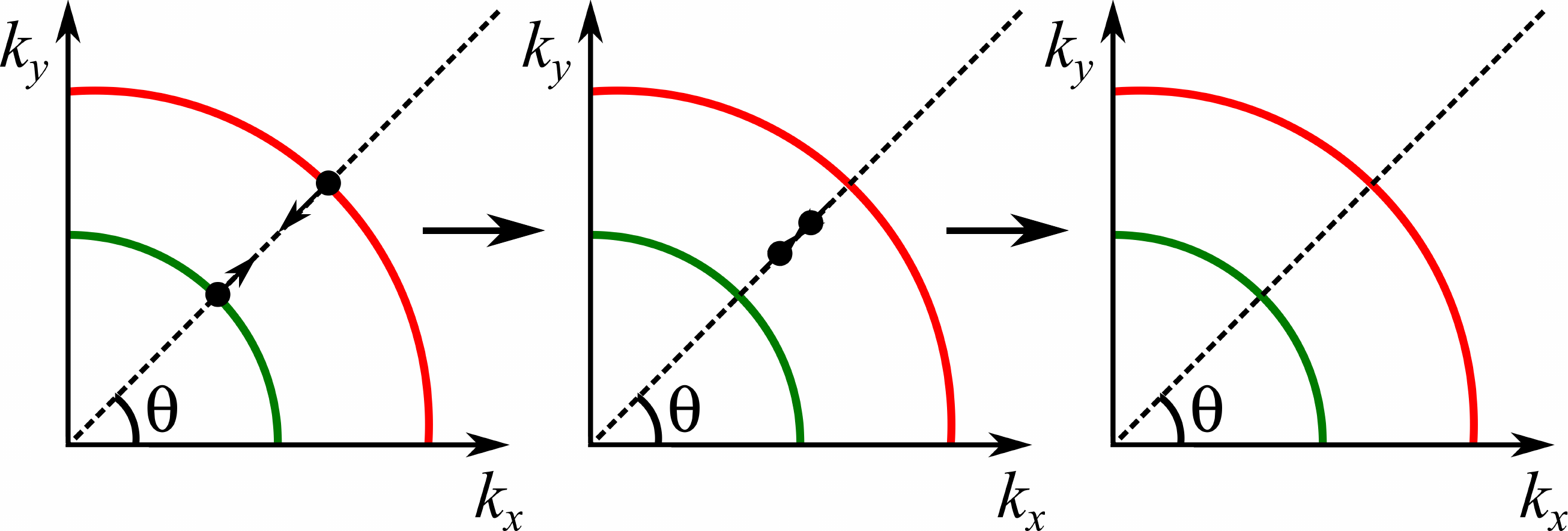}}
\centering{}\caption{The nodal points (black dots) in a $d$-wave FeSC with FSs made out of $d_{xz}$ and $d_{yz}$ orbitals.  Red and green lines are the two FSs from the normal state. When the magnitude of the $d$-wave gap $\Delta$ increases, nodal points move along the BZ diagonal, and  merge and disappear at some $\Delta_{\mathrm{crit}}.$ }
\label{sk3}
\end{figure}

We next consider the model of FeSC with the $d$-wave gap structure \cite{Chubukov2016,Nica2017}. The model is for a heavily hole doped FeSC  with
two $\Gamma-$ centered hole pockets and no electron pockets.
The hole pockets are made out of  $d_{xz}$ and $d_{yz}$ orbitals, and orbital content is rotated by $90^o$ between the two pockets.
Because the orbital content varies along the Fermi surfaces, the interactions in the band basis are angle-dependent and have both $s$-wave and $d$-wave components. We assume that $d$-wave interaction is attractive and stronger than $s$-wave one, such that the system develops   $d_{x^2-y^2}$ superconductivity below a certain $T$.

The $d$-wave gap equation in the band basis has been analyzed in~\cite{Chubukov2016,Nica2017}.  The kinetic energy is
\begin{equation}
H_0=\sum_{k,\alpha} \left(\epsilon_{1,k} c_{1,k}^{\dagger} c_{1,k} + \epsilon_{2,k} c_{2,k}^{\dagger} c_{2,k} \right),
\end{equation}
 where $\epsilon_{1,2,k}=\mu-k^2/(2m_{1,2})$ and we consider $m_1 \neq m_2$.
By symmetry, the pairing interaction couples intra-pocket pairing condensates $\av{c_{1,k\alpha}^{\dagger} c_{1,-k\beta}^{\dagger}}$ and  $\av{c_{2,k\alpha}^{\dagger} c_{2,-k\beta}}$, and inter-pocket pairing condensates $\av{c_{1,k\alpha}^{\dagger} c_{2,-k\beta}^{\dagger}}$  and $\av{c_{2,k\alpha}^{\dagger} c_{1,-k\beta}}$.
For the case  when the interaction in the band basis is obtained from a local Hubbard-Hund interaction in the orbital basis,
 the anomalous part of the BCS Hamiltonian is
\begin{equation}
\begin{split}
H_{\Delta}=\Delta_a \sum_k i \sigma^y_{\alpha\beta} \left(c_{1,k\alpha}^{\dagger} c_{1,-k\beta}^{\dagger} - c_{2,k\alpha}^{\dagger} c_{2,-k\beta}  \right) + \\
+\Delta_b \sum_k i \sigma^y_{\alpha\beta} \left(c_{1,k\alpha}^{\dagger} c_{2,-k\beta}^{\dagger} + c_{2,k\alpha}^{\dagger} c_{1,-k\beta}  \right) + \mathrm{H.c.}
\end{split}
\end{equation}
where $\Delta_a=\Delta \,  \mathrm{cos} 2\theta$ and $\Delta_b=\Delta \, \mathrm{sin} 2\theta$.

Diagonalizing this BCS  Hamiltonian, we obtain two bands, $a$ and $b$, with the dispersion
\begin{equation}
E_{a,b}(k)=\sqrt{\Delta^2 \mathrm{cos}^2 (2\theta) + \epsilon^2_{a,b}(k)},
\label{dwavedisp}
\end{equation}
where
\begin{equation}
\epsilon_{a,b}(k)=\mathrm{sgn}(\epsilon_{1,k}+\epsilon_{2,k})\sqrt{\left( \frac{\epsilon_{1,k}+\epsilon_{2,k}}{2} \right)^2 + \Delta^2 \mathrm{sin}^2 (2\theta)} \; \pm \frac{\epsilon_{1,k}-\epsilon_{2,k}}{2}.
\end{equation}

When the two Fermi surfaces are far apart, $\epsilon_a \approx \epsilon_{1,k}$ and $\epsilon_b \approx \epsilon_{2,k}$. In this limit, we have a conventional $d$-wave gap structure with nodal points on each Fermi surface, along the diagonals.  However, when $\Delta$ is comparable to the energy difference between $\epsilon_1$ and $\epsilon_2$, when one of $\epsilon$ vanishes, the nodal points move away from the two Fermi surfaces into the region between them (see Fig. \ref{sk3}).  At some critical $\Delta$, the two nodal points along each diagonal merge and disappear, leaving a $d$-wave superconductor nodeless.

\subsection{The winding number}

Without loss of generality we set $m_2>m_1$. Inside the smaller Fermi surface  $\epsilon_{1,k}<0$ and $\epsilon_{2,k} <0$.  Upon crossing the smaller Fermi surface
 $\epsilon_{1,k}$ changes sign, but $\epsilon_{2,k}$ remains negative, i.e.,  $\mathrm{sgn}(\epsilon_{1,k}+\epsilon_{2,k}) =-1$.
Near the larger Fermi surface $\epsilon_{2,k}\simeq0$ and $\epsilon_{1,k} >0$.  Then $\mathrm{sgn}(\epsilon_{1,k}+\epsilon_{2,k}) = 1$. As a result, near each of the two  nodal points  $\epsilon_a = \xi = - \epsilon_b$.
Using this, we  construct effective Dirac Hamiltonians $H_a$ and $H_b$:
\begin{equation}
H_{a,b} = \begin{pmatrix}
\pm\xi & 2\Delta (\theta - \pi/4) \\
2\Delta (\theta - \pi/4) & \mp\xi
\end{pmatrix}.
\label{Hab}
\end{equation}
The corresponding matrices $A$ are
\begin{equation}
A_{a,b} = \begin{pmatrix}
 2\Delta & 0\\
0& \pm \frac{d\xi}{dk}
\end{pmatrix}
\end{equation}
Then ${\text sign} [\mathrm{det}(A_a)] = - {\text sign} [\mathrm{det}(A_b)]$, i.e., the two nodal points along each diagonal have opposite winding numbers.

\subsection{Numerical analysis}

\begin{figure}[t]
\begin{minipage}[h]{0.49\linewidth}
\center{\includegraphics[width=0.99\linewidth]{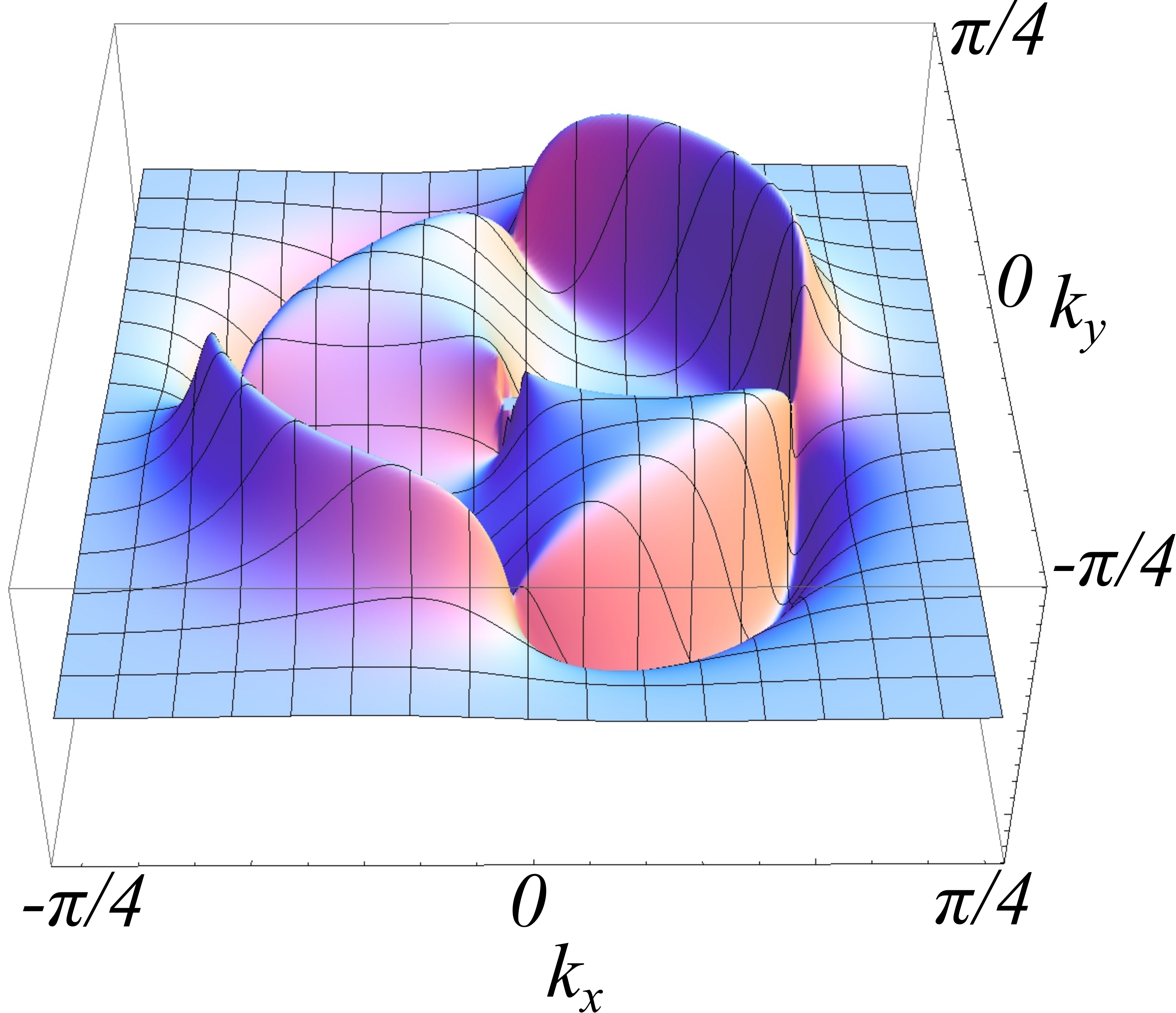} \\ $\mathit{a}$-band}
\end{minipage}
\hfill
\begin{minipage}[h]{0.49\linewidth}
\center{\includegraphics[width=0.99\linewidth]{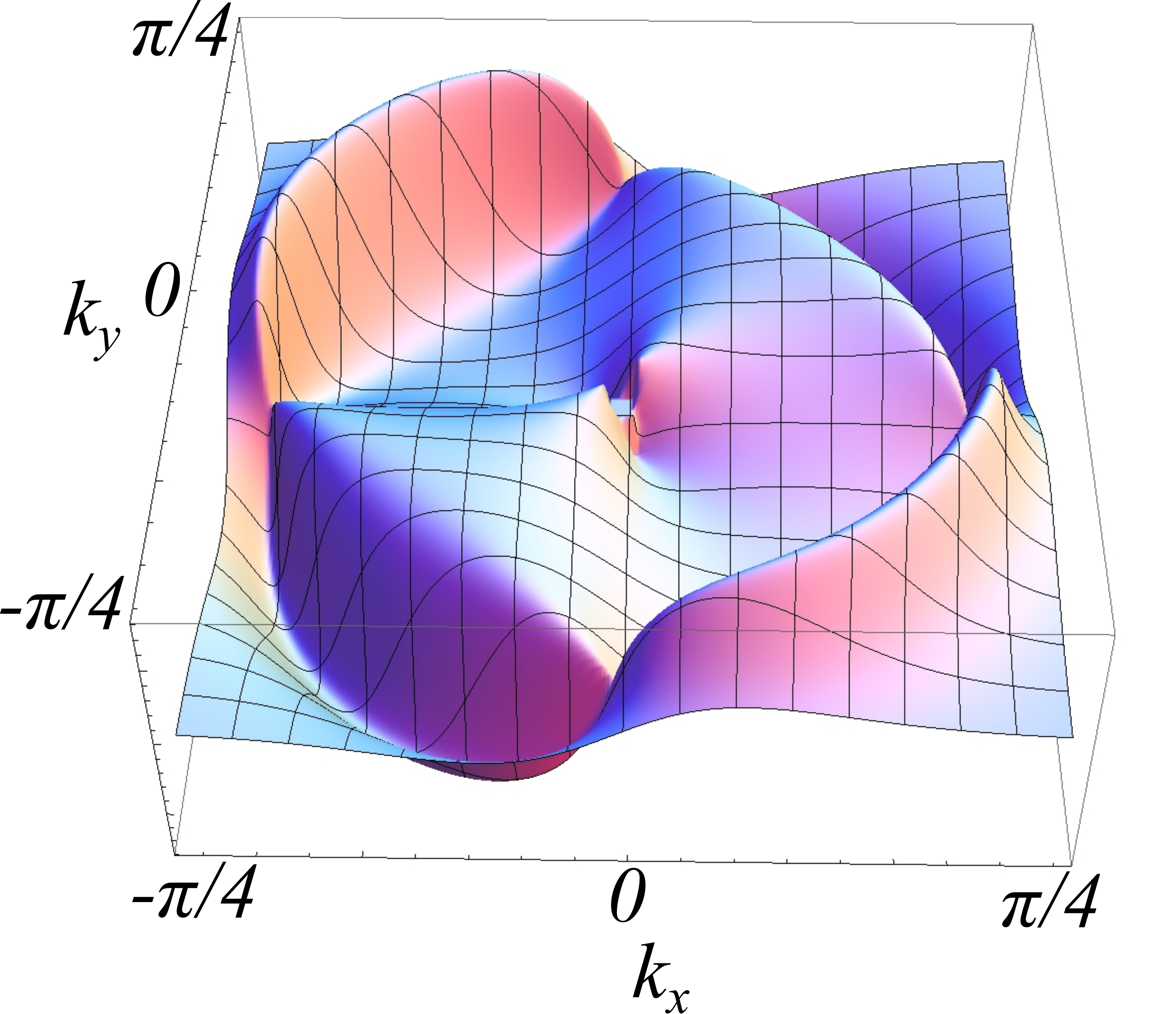} \\ $\mathit{b}$-band}
\end{minipage}
\centering{}\caption{The grid Berry curvature $\tilde{B}$ computed for the two bands $a$ and $b$ (each with nodal points) for a model of a $d$-wave superconductor, Eq. (\ref{dwavedisp}).
  We set $\Delta=0.3, \mu=1, \alpha=1.$ We see that the sign of the Berry curvature changes between the two nodal points along the same diagonal.}
\label{d_phase}
\end{figure}

We computed the Berry curvature separately for the effective Hamiltonians
$H_{a}$ and $H_b$. The results are shown in Fig. \ref{d_phase}. We see that the Berry curvature has opposite signs
for a  pair of nodal points along the same diagonal, hence these points have the winding numbers of opposite sign. This agrees with the  analytic result.

\section{Conclusions}
\label{sec_6}

In this paper we analyzed the merging and disappearance of the nodal points in FeSCs  from  topological perspective.
We considered two models with  different pairing symmetry -- $\mathit{s}$-wave ($s^{\pm}$) and $\mathit{d}$-wave.
For an $s^{+-}$-wave superconductor we considered the model with accidental nodes on the two electron pockets. We manipulated the position of the nodes by varying the degree of hybridization between the two electron pockets.  We considered first the special case when the electron pockets are  circular, and then a  generic case when they are elliptical.  In both cases increasing the strength of hybridization gives rise to the Lifshitz transition in which neighboring nodal points merge and annihilate. For the case of circular pockets we showed that of eight nodal points four have positive winding number $Q=+1$ and four have  $Q=-1$.  We showed that the nodal points, which merge at the Lifshitz transition,  have opposite winding numbers.
In the case of elliptical pockets,  we focused on the case when, upon the increase of hybridization, first eight new nodal points are created in pairs, and then new nodal points merge with the existing ones. We showed that in each pair the two  emerging nodes have opposite signs of the winding number.
And the winding numbers of the newly created and the existing nodal points, which  merge and annihilate at larger hybridization, are also opposite.
  As a result, the  net topological invariant is conserved in the Lifshitz transition and  from this perspective the transition from a nodal to full gap $s^{+-}$ superconductor   can be labeled as non-topological one.

For $\mathit{d}$-wave gap symmetry we considered a model with two hole pockets made out of $d_{xz}$ and $d_{yz}$ orbitals.
The pairing condensate in this model necessary contains intra-pocket and inter-pocket components.  The latter move the nodal points away from the Fermi surfaces, into the area in between the pockets.  As the pairing gap increases (or the distance between the pockets decreases), the two nodal points along each diagonal come closer to each other and eventually merge and disappear via Lifshitz transition.  We showed that the winding numbers of these nodal points are again  $Q = \pm 1$. Then the
net winding number is zero, and  the Lifshitz transition in a $d$-wave case also  can be labeled as non-topological.

The merging and annihilation of nodal points  has been well studied  in Dirac and Weyl semi-metals  which undergo a transition into an insulator under a variation of certain system parameters~\cite{Vafek2014}. Several authors have shown that in a semi-metal-to-insulator transition, the merging nodal points have opposite winding numbers \cite{Murakami2008,Murakami2007}.
We demonstrated  that the same is true in nodal-to-full gap transitions in $s$-wave and $d$-wave FeSCs.

\section{Acknowledgments}
 We thank A. Hinojosa, D. Shaffer, and O. Vafek for useful discussions.  The work was supported by the Office of Basic Energy Sciences U. S. Department of Energy
under the award DE-SC0014402.

\bibliography{biblio}

\begin{thebibliography}{31}%
\makeatletter
\providecommand \@ifxundefined [1]{%
 \@ifx{#1\undefined}
}%
\providecommand \@ifnum [1]{%
 \ifnum #1\expandafter \@firstoftwo
 \else \expandafter \@secondoftwo
 \fi
}%
\providecommand \@ifx [1]{%
 \ifx #1\expandafter \@firstoftwo
 \else \expandafter \@secondoftwo
 \fi
}%
\providecommand \natexlab [1]{#1}%
\providecommand \enquote  [1]{``#1''}%
\providecommand \bibnamefont  [1]{#1}%
\providecommand \bibfnamefont [1]{#1}%
\providecommand \citenamefont [1]{#1}%
\providecommand \href@noop [0]{\@secondoftwo}%
\providecommand \href [0]{\begingroup \@sanitize@url \@href}%
\providecommand \@href[1]{\@@startlink{#1}\@@href}%
\providecommand \@@href[1]{\endgroup#1\@@endlink}%
\providecommand \@sanitize@url [0]{\catcode `\\12\catcode `\$12\catcode
  `\&12\catcode `\#12\catcode `\^12\catcode `\_12\catcode `\%12\relax}%
\providecommand \@@startlink[1]{}%
\providecommand \@@endlink[0]{}%
\providecommand \url  [0]{\begingroup\@sanitize@url \@url }%
\providecommand \@url [1]{\endgroup\@href {#1}{\urlprefix }}%
\providecommand \urlprefix  [0]{URL }%
\providecommand \Eprint [0]{\href }%
\providecommand \doibase [0]{http://dx.doi.org/}%
\providecommand \selectlanguage [0]{\@gobble}%
\providecommand \bibinfo  [0]{\@secondoftwo}%
\providecommand \bibfield  [0]{\@secondoftwo}%
\providecommand \translation [1]{[#1]}%
\providecommand \BibitemOpen [0]{}%
\providecommand \bibitemStop [0]{}%
\providecommand \bibitemNoStop [0]{.\EOS\space}%
\providecommand \EOS [0]{\spacefactor3000\relax}%
\providecommand \BibitemShut  [1]{\csname bibitem#1\endcsname}%
\let\auto@bib@innerbib\@empty
\bibitem [{\citenamefont {Thouless}\ \emph {et~al.}(1982)\citenamefont
  {Thouless}, \citenamefont {Kohmoto}, \citenamefont {Nightingale},\ and\
  \citenamefont {den Nijs}}]{Thouless1982}%
  \BibitemOpen
  \bibfield  {author} {\bibinfo {author} {\bibfnamefont {D.~J.}\ \bibnamefont
  {Thouless}}, \bibinfo {author} {\bibfnamefont {M.}~\bibnamefont {Kohmoto}},
  \bibinfo {author} {\bibfnamefont {M.~P.}\ \bibnamefont {Nightingale}}, \ and\
  \bibinfo {author} {\bibfnamefont {M.}~\bibnamefont {den Nijs}},\ }\href
  {\doibase 10.1103/PhysRevLett.49.405} {\bibfield  {journal} {\bibinfo
  {journal} {Phys. Rev. Lett.}\ }\textbf {\bibinfo {volume} {49}},\ \bibinfo
  {pages} {405} (\bibinfo {year} {1982})}\BibitemShut {NoStop}%
\bibitem [{\citenamefont {Simon}(1983)}]{Simon1983}%
  \BibitemOpen
  \bibfield  {author} {\bibinfo {author} {\bibfnamefont {B.}~\bibnamefont
  {Simon}},\ }\href {\doibase 10.1103/PhysRevLett.51.2167} {\bibfield
  {journal} {\bibinfo  {journal} {Phys. Rev. Lett.}\ }\textbf {\bibinfo
  {volume} {51}},\ \bibinfo {pages} {2167} (\bibinfo {year}
  {1983})}\BibitemShut {NoStop}%
\bibitem [{\citenamefont {Berry}(1984)}]{Berry1984}%
  \BibitemOpen
  \bibfield  {author} {\bibinfo {author} {\bibfnamefont {M.}~\bibnamefont
  {Berry}},\ }\href {\doibase 10.1098/rspa.1984.0023} {\bibfield  {journal}
  {\bibinfo  {journal} {Proceedings of the Royal Society of London A:
  Mathematical, Physical and Engineering Sciences}\ }\textbf {\bibinfo {volume}
  {392}},\ \bibinfo {pages} {45} (\bibinfo {year} {1984})},\ \Eprint
  {http://arxiv.org/abs/http://rspa.royalsocietypublishing.org/content/392/1802/45.full.pdf}
  {http://rspa.royalsocietypublishing.org/content/392/1802/45.full.pdf}
  \BibitemShut {NoStop}%
\bibitem [{\citenamefont {Bernevig}\ and\ \citenamefont
  {Zhang}(2006)}]{Bernevig2006}%
  \BibitemOpen
  \bibfield  {author} {\bibinfo {author} {\bibfnamefont {B.~A.}\ \bibnamefont
  {Bernevig}}\ and\ \bibinfo {author} {\bibfnamefont {S.-C.}\ \bibnamefont
  {Zhang}},\ }\href {\doibase 10.1103/PhysRevLett.96.106802} {\bibfield
  {journal} {\bibinfo  {journal} {Phys. Rev. Lett.}\ }\textbf {\bibinfo
  {volume} {96}},\ \bibinfo {pages} {106802} (\bibinfo {year}
  {2006})}\BibitemShut {NoStop}%
\bibitem [{\citenamefont {Bernevig}\ \emph {et~al.}(2006)\citenamefont
  {Bernevig}, \citenamefont {Hughes},\ and\ \citenamefont
  {Zhang}}]{Bernevig2006Science}%
  \BibitemOpen
  \bibfield  {author} {\bibinfo {author} {\bibfnamefont {B.~A.}\ \bibnamefont
  {Bernevig}}, \bibinfo {author} {\bibfnamefont {T.~L.}\ \bibnamefont
  {Hughes}}, \ and\ \bibinfo {author} {\bibfnamefont {S.-C.}\ \bibnamefont
  {Zhang}},\ }\href {\doibase 10.1126/science.1133734} {\bibfield  {journal}
  {\bibinfo  {journal} {Science}\ }\textbf {\bibinfo {volume} {314}},\ \bibinfo
  {pages} {1757} (\bibinfo {year} {2006})},\ \Eprint
  {http://arxiv.org/abs/http://science.sciencemag.org/content/314/5806/1757.full.pdf}
  {http://science.sciencemag.org/content/314/5806/1757.full.pdf} \BibitemShut
  {NoStop}%
\bibitem [{\citenamefont {Wan}\ \emph {et~al.}(2011)\citenamefont {Wan},
  \citenamefont {Turner}, \citenamefont {Vishwanath},\ and\ \citenamefont
  {Savrasov}}]{Wan2011}%
  \BibitemOpen
  \bibfield  {author} {\bibinfo {author} {\bibfnamefont {X.}~\bibnamefont
  {Wan}}, \bibinfo {author} {\bibfnamefont {A.~M.}\ \bibnamefont {Turner}},
  \bibinfo {author} {\bibfnamefont {A.}~\bibnamefont {Vishwanath}}, \ and\
  \bibinfo {author} {\bibfnamefont {S.~Y.}\ \bibnamefont {Savrasov}},\ }\href
  {\doibase 10.1103/PhysRevB.83.205101} {\bibfield  {journal} {\bibinfo
  {journal} {Phys. Rev. B}\ }\textbf {\bibinfo {volume} {83}},\ \bibinfo
  {pages} {205101} (\bibinfo {year} {2011})}\BibitemShut {NoStop}%
\bibitem [{\citenamefont {Xu}\ \emph {et~al.}(2015)\citenamefont {Xu},
  \citenamefont {Belopolski}, \citenamefont {Alidoust}, \citenamefont
  {Neupane}, \citenamefont {Bian}, \citenamefont {Zhang}, \citenamefont
  {Sankar}, \citenamefont {Chang}, \citenamefont {Yuan}, \citenamefont {Lee},
  \citenamefont {Huang}, \citenamefont {Zheng}, \citenamefont {Ma},
  \citenamefont {Sanchez}, \citenamefont {Wang}, \citenamefont {Bansil},
  \citenamefont {Chou}, \citenamefont {Shibayev}, \citenamefont {Lin},
  \citenamefont {Jia},\ and\ \citenamefont {Hasan}}]{Xu2015}%
  \BibitemOpen
  \bibfield  {author} {\bibinfo {author} {\bibfnamefont {S.-Y.}\ \bibnamefont
  {Xu}}, \bibinfo {author} {\bibfnamefont {I.}~\bibnamefont {Belopolski}},
  \bibinfo {author} {\bibfnamefont {N.}~\bibnamefont {Alidoust}}, \bibinfo
  {author} {\bibfnamefont {M.}~\bibnamefont {Neupane}}, \bibinfo {author}
  {\bibfnamefont {G.}~\bibnamefont {Bian}}, \bibinfo {author} {\bibfnamefont
  {C.}~\bibnamefont {Zhang}}, \bibinfo {author} {\bibfnamefont
  {R.}~\bibnamefont {Sankar}}, \bibinfo {author} {\bibfnamefont
  {G.}~\bibnamefont {Chang}}, \bibinfo {author} {\bibfnamefont
  {Z.}~\bibnamefont {Yuan}}, \bibinfo {author} {\bibfnamefont {C.-C.}\
  \bibnamefont {Lee}}, \bibinfo {author} {\bibfnamefont {S.-M.}\ \bibnamefont
  {Huang}}, \bibinfo {author} {\bibfnamefont {H.}~\bibnamefont {Zheng}},
  \bibinfo {author} {\bibfnamefont {J.}~\bibnamefont {Ma}}, \bibinfo {author}
  {\bibfnamefont {D.~S.}\ \bibnamefont {Sanchez}}, \bibinfo {author}
  {\bibfnamefont {B.}~\bibnamefont {Wang}}, \bibinfo {author} {\bibfnamefont
  {A.}~\bibnamefont {Bansil}}, \bibinfo {author} {\bibfnamefont
  {F.}~\bibnamefont {Chou}}, \bibinfo {author} {\bibfnamefont {P.~P.}\
  \bibnamefont {Shibayev}}, \bibinfo {author} {\bibfnamefont {H.}~\bibnamefont
  {Lin}}, \bibinfo {author} {\bibfnamefont {S.}~\bibnamefont {Jia}}, \ and\
  \bibinfo {author} {\bibfnamefont {M.~Z.}\ \bibnamefont {Hasan}},\ }\href
  {\doibase 10.1126/science.aaa9297} {\bibfield  {journal} {\bibinfo  {journal}
  {Science}\ }\textbf {\bibinfo {volume} {349}},\ \bibinfo {pages} {613}
  (\bibinfo {year} {2015})},\ \Eprint
  {http://arxiv.org/abs/http://science.sciencemag.org/content/349/6248/613.full.pdf}
  {http://science.sciencemag.org/content/349/6248/613.full.pdf} \BibitemShut
  {NoStop}%
\bibitem [{\citenamefont {Kamihara}\ \emph {et~al.}(2008)\citenamefont
  {Kamihara}, \citenamefont {Watanabe}, \citenamefont {Hirano},\ and\
  \citenamefont {Hosono}}]{Kamihara2008}%
  \BibitemOpen
  \bibfield  {author} {\bibinfo {author} {\bibfnamefont {Y.}~\bibnamefont
  {Kamihara}}, \bibinfo {author} {\bibfnamefont {T.}~\bibnamefont {Watanabe}},
  \bibinfo {author} {\bibfnamefont {M.}~\bibnamefont {Hirano}}, \ and\ \bibinfo
  {author} {\bibfnamefont {H.}~\bibnamefont {Hosono}},\ }\href {\doibase
  10.1021/ja800073m} {\bibfield  {journal} {\bibinfo  {journal} {Journal of the
  American Chemical Society}\ }\textbf {\bibinfo {volume} {130}},\ \bibinfo
  {pages} {3296} (\bibinfo {year} {2008})},\ \bibinfo {note} {pMID: 18293989},\
  \Eprint {http://arxiv.org/abs/http://dx.doi.org/10.1021/ja800073m}
  {http://dx.doi.org/10.1021/ja800073m} \BibitemShut {NoStop}%
\bibitem [{\citenamefont {de~la Cruz}\ \emph {et~al.}(2008)\citenamefont {de~la
  Cruz}, \citenamefont {Huang}, \citenamefont {Lynn}, \citenamefont {Li},
  \citenamefont {II}, \citenamefont {Zarestky}, \citenamefont {Mook},
  \citenamefont {Chen}, \citenamefont {Luo}, \citenamefont {Wang} \emph
  {et~al.}}]{delaCruz2008}%
  \BibitemOpen
  \bibfield  {author} {\bibinfo {author} {\bibfnamefont {C.}~\bibnamefont
  {de~la Cruz}}, \bibinfo {author} {\bibfnamefont {Q.}~\bibnamefont {Huang}},
  \bibinfo {author} {\bibfnamefont {J.}~\bibnamefont {Lynn}}, \bibinfo {author}
  {\bibfnamefont {J.}~\bibnamefont {Li}}, \bibinfo {author} {\bibfnamefont
  {W.~R.}\ \bibnamefont {II}}, \bibinfo {author} {\bibfnamefont
  {J.}~\bibnamefont {Zarestky}}, \bibinfo {author} {\bibfnamefont
  {H.}~\bibnamefont {Mook}}, \bibinfo {author} {\bibfnamefont {G.}~\bibnamefont
  {Chen}}, \bibinfo {author} {\bibfnamefont {J.}~\bibnamefont {Luo}}, \bibinfo
  {author} {\bibfnamefont {N.}~\bibnamefont {Wang}},  \emph {et~al.},\ }\href
  {http://dx.doi.org/10.1038/nature07057} {\bibfield  {journal} {\bibinfo
  {journal} {Nature}\ }\textbf {\bibinfo {volume} {453}},\ \bibinfo {pages}
  {899} (\bibinfo {year} {2008})}\BibitemShut {NoStop}%
\bibitem [{\citenamefont {Liu}\ \emph {et~al.}(2010)\citenamefont {Liu},
  \citenamefont {Kondo}, \citenamefont {Fernandes}, \citenamefont {Palczewski},
  \citenamefont {Mun}, \citenamefont {Ni}, \citenamefont {Thaler},
  \citenamefont {Bostwick}, \citenamefont {Rotenberg}, \citenamefont
  {Schmalian} \emph {et~al.}}]{Liu2010}%
  \BibitemOpen
  \bibfield  {author} {\bibinfo {author} {\bibfnamefont {C.}~\bibnamefont
  {Liu}}, \bibinfo {author} {\bibfnamefont {T.}~\bibnamefont {Kondo}}, \bibinfo
  {author} {\bibfnamefont {R.~M.}\ \bibnamefont {Fernandes}}, \bibinfo {author}
  {\bibfnamefont {A.~D.}\ \bibnamefont {Palczewski}}, \bibinfo {author}
  {\bibfnamefont {E.~D.}\ \bibnamefont {Mun}}, \bibinfo {author} {\bibfnamefont
  {N.}~\bibnamefont {Ni}}, \bibinfo {author} {\bibfnamefont {A.~N.}\
  \bibnamefont {Thaler}}, \bibinfo {author} {\bibfnamefont {A.}~\bibnamefont
  {Bostwick}}, \bibinfo {author} {\bibfnamefont {E.}~\bibnamefont {Rotenberg}},
  \bibinfo {author} {\bibfnamefont {J.}~\bibnamefont {Schmalian}},  \emph
  {et~al.},\ }\href@noop {} {\bibfield  {journal} {\bibinfo  {journal} {Nature
  Physics}\ }\textbf {\bibinfo {volume} {6}},\ \bibinfo {pages} {419} (\bibinfo
  {year} {2010})}\BibitemShut {NoStop}%
\bibitem [{\citenamefont {Putti}\ \emph {et~al.}(2010)\citenamefont {Putti},
  \citenamefont {Pallecchi}, \citenamefont {Bellingeri}, \citenamefont
  {Cimberle}, \citenamefont {Tropeano}, \citenamefont {Ferdeghini},
  \citenamefont {Palenzona}, \citenamefont {Tarantini}, \citenamefont
  {Yamamoto}, \citenamefont {Jiang} \emph {et~al.}}]{Putti2010}%
  \BibitemOpen
  \bibfield  {author} {\bibinfo {author} {\bibfnamefont {M.}~\bibnamefont
  {Putti}}, \bibinfo {author} {\bibfnamefont {I.}~\bibnamefont {Pallecchi}},
  \bibinfo {author} {\bibfnamefont {E.}~\bibnamefont {Bellingeri}}, \bibinfo
  {author} {\bibfnamefont {M.}~\bibnamefont {Cimberle}}, \bibinfo {author}
  {\bibfnamefont {M.}~\bibnamefont {Tropeano}}, \bibinfo {author}
  {\bibfnamefont {C.}~\bibnamefont {Ferdeghini}}, \bibinfo {author}
  {\bibfnamefont {A.}~\bibnamefont {Palenzona}}, \bibinfo {author}
  {\bibfnamefont {C.}~\bibnamefont {Tarantini}}, \bibinfo {author}
  {\bibfnamefont {A.}~\bibnamefont {Yamamoto}}, \bibinfo {author}
  {\bibfnamefont {J.}~\bibnamefont {Jiang}},  \emph {et~al.},\ }\href@noop {}
  {\bibfield  {journal} {\bibinfo  {journal} {Superconductor Science and
  Technology}\ }\textbf {\bibinfo {volume} {23}},\ \bibinfo {pages} {034003}
  (\bibinfo {year} {2010})}\BibitemShut {NoStop}%
\bibitem [{\citenamefont {Hanaguri}\ \emph {et~al.}(2010)\citenamefont
  {Hanaguri}, \citenamefont {Niitaka}, \citenamefont {Kuroki},\ and\
  \citenamefont {Takagi}}]{Hanaguri2010}%
  \BibitemOpen
  \bibfield  {author} {\bibinfo {author} {\bibfnamefont {T.}~\bibnamefont
  {Hanaguri}}, \bibinfo {author} {\bibfnamefont {S.}~\bibnamefont {Niitaka}},
  \bibinfo {author} {\bibfnamefont {K.}~\bibnamefont {Kuroki}}, \ and\ \bibinfo
  {author} {\bibfnamefont {H.}~\bibnamefont {Takagi}},\ }\href {\doibase
  10.1126/science.1187399} {\bibfield  {journal} {\bibinfo  {journal}
  {Science}\ }\textbf {\bibinfo {volume} {328}},\ \bibinfo {pages} {474}
  (\bibinfo {year} {2010})},\ \Eprint
  {http://arxiv.org/abs/http://science.sciencemag.org/content/328/5977/474.full.pdf}
  {http://science.sciencemag.org/content/328/5977/474.full.pdf} \BibitemShut
  {NoStop}%
\bibitem [{\citenamefont {Maier}\ \emph {et~al.}(2011)\citenamefont {Maier},
  \citenamefont {Graser}, \citenamefont {Hirschfeld},\ and\ \citenamefont
  {Scalapino}}]{Maier2011}%
  \BibitemOpen
  \bibfield  {author} {\bibinfo {author} {\bibfnamefont {T.~A.}\ \bibnamefont
  {Maier}}, \bibinfo {author} {\bibfnamefont {S.}~\bibnamefont {Graser}},
  \bibinfo {author} {\bibfnamefont {P.~J.}\ \bibnamefont {Hirschfeld}}, \ and\
  \bibinfo {author} {\bibfnamefont {D.~J.}\ \bibnamefont {Scalapino}},\ }\href
  {\doibase 10.1103/PhysRevB.83.100515} {\bibfield  {journal} {\bibinfo
  {journal} {Phys. Rev. B}\ }\textbf {\bibinfo {volume} {83}},\ \bibinfo
  {pages} {100515} (\bibinfo {year} {2011})}\BibitemShut {NoStop}%
\bibitem [{\citenamefont {Hirschfeld}\ \emph {et~al.}(2011)\citenamefont
  {Hirschfeld}, \citenamefont {Korshunov},\ and\ \citenamefont
  {Mazin}}]{Hirschfeld2011}%
  \BibitemOpen
  \bibfield  {author} {\bibinfo {author} {\bibfnamefont {P.}~\bibnamefont
  {Hirschfeld}}, \bibinfo {author} {\bibfnamefont {M.}~\bibnamefont
  {Korshunov}}, \ and\ \bibinfo {author} {\bibfnamefont {I.}~\bibnamefont
  {Mazin}},\ }\href@noop {} {\bibfield  {journal} {\bibinfo  {journal} {Reports
  on Progress in Physics}\ }\textbf {\bibinfo {volume} {74}},\ \bibinfo {pages}
  {124508} (\bibinfo {year} {2011})}\BibitemShut {NoStop}%
\bibitem [{\citenamefont {Khodas}\ and\ \citenamefont
  {Chubukov}(2012)}]{Khodas2012}%
  \BibitemOpen
  \bibfield  {author} {\bibinfo {author} {\bibfnamefont {M.}~\bibnamefont
  {Khodas}}\ and\ \bibinfo {author} {\bibfnamefont {A.~V.}\ \bibnamefont
  {Chubukov}},\ }\href {\doibase 10.1103/PhysRevLett.108.247003} {\bibfield
  {journal} {\bibinfo  {journal} {Phys. Rev. Lett.}\ }\textbf {\bibinfo
  {volume} {108}},\ \bibinfo {pages} {247003} (\bibinfo {year}
  {2012})}\BibitemShut {NoStop}%
\bibitem [{\citenamefont {Chubukov}\ \emph {et~al.}(2016)\citenamefont
  {Chubukov}, \citenamefont {Vafek},\ and\ \citenamefont
  {Fernandes}}]{Chubukov2016}%
  \BibitemOpen
  \bibfield  {author} {\bibinfo {author} {\bibfnamefont {A.~V.}\ \bibnamefont
  {Chubukov}}, \bibinfo {author} {\bibfnamefont {O.}~\bibnamefont {Vafek}}, \
  and\ \bibinfo {author} {\bibfnamefont {R.~M.}\ \bibnamefont {Fernandes}},\
  }\href {\doibase 10.1103/PhysRevB.94.174518} {\bibfield  {journal} {\bibinfo
  {journal} {Phys. Rev. B}\ }\textbf {\bibinfo {volume} {94}},\ \bibinfo
  {pages} {174518} (\bibinfo {year} {2016})}\BibitemShut {NoStop}%
\bibitem [{\citenamefont {Lifshitz}(1960)}]{Lifshitz1960}%
  \BibitemOpen
  \bibfield  {author} {\bibinfo {author} {\bibfnamefont {I.~M.}\ \bibnamefont
  {Lifshitz}},\ }\href@noop {} {\bibfield  {journal} {\bibinfo  {journal} {Sov.
  Phys. JETP}\ }\textbf {\bibinfo {volume} {11}},\ \bibinfo {pages} {1130}
  (\bibinfo {year} {1960})}\BibitemShut {NoStop}%
\bibitem [{\citenamefont {Liu}\ \emph {et~al.}(2011)\citenamefont {Liu},
  \citenamefont {Palczewski}, \citenamefont {Dhaka}, \citenamefont {Kondo},
  \citenamefont {Fernandes}, \citenamefont {Mun}, \citenamefont {Hodovanets},
  \citenamefont {Thaler}, \citenamefont {Schmalian}, \citenamefont {Bud'ko},
  \citenamefont {Canfield},\ and\ \citenamefont {Kaminski}}]{Liu2011}%
  \BibitemOpen
  \bibfield  {author} {\bibinfo {author} {\bibfnamefont {C.}~\bibnamefont
  {Liu}}, \bibinfo {author} {\bibfnamefont {A.~D.}\ \bibnamefont {Palczewski}},
  \bibinfo {author} {\bibfnamefont {R.~S.}\ \bibnamefont {Dhaka}}, \bibinfo
  {author} {\bibfnamefont {T.}~\bibnamefont {Kondo}}, \bibinfo {author}
  {\bibfnamefont {R.~M.}\ \bibnamefont {Fernandes}}, \bibinfo {author}
  {\bibfnamefont {E.~D.}\ \bibnamefont {Mun}}, \bibinfo {author} {\bibfnamefont
  {H.}~\bibnamefont {Hodovanets}}, \bibinfo {author} {\bibfnamefont {A.~N.}\
  \bibnamefont {Thaler}}, \bibinfo {author} {\bibfnamefont {J.}~\bibnamefont
  {Schmalian}}, \bibinfo {author} {\bibfnamefont {S.~L.}\ \bibnamefont
  {Bud'ko}}, \bibinfo {author} {\bibfnamefont {P.~C.}\ \bibnamefont
  {Canfield}}, \ and\ \bibinfo {author} {\bibfnamefont {A.}~\bibnamefont
  {Kaminski}},\ }\href {\doibase 10.1103/PhysRevB.84.020509} {\bibfield
  {journal} {\bibinfo  {journal} {Phys. Rev. B}\ }\textbf {\bibinfo {volume}
  {84}},\ \bibinfo {pages} {020509} (\bibinfo {year} {2011})}\BibitemShut
  {NoStop}%
\bibitem [{\citenamefont {Hinojosa}\ and\ \citenamefont
  {Chubukov}(2015)}]{Hinojosa2015}%
  \BibitemOpen
  \bibfield  {author} {\bibinfo {author} {\bibfnamefont {A.}~\bibnamefont
  {Hinojosa}}\ and\ \bibinfo {author} {\bibfnamefont {A.~V.}\ \bibnamefont
  {Chubukov}},\ }\href {\doibase 10.1103/PhysRevB.91.224502} {\bibfield
  {journal} {\bibinfo  {journal} {Phys. Rev. B}\ }\textbf {\bibinfo {volume}
  {91}},\ \bibinfo {pages} {224502} (\bibinfo {year} {2015})}\BibitemShut
  {NoStop}%
\bibitem [{\citenamefont {Nica}\ \emph {et~al.}(2017)\citenamefont {Nica},
  \citenamefont {Yu},\ and\ \citenamefont {Si}}]{Nica2017}%
  \BibitemOpen
  \bibfield  {author} {\bibinfo {author} {\bibfnamefont {E.~M.}\ \bibnamefont
  {Nica}}, \bibinfo {author} {\bibfnamefont {R.}~\bibnamefont {Yu}}, \ and\
  \bibinfo {author} {\bibfnamefont {Q.}~\bibnamefont {Si}},\ }\href@noop {}
  {\bibfield  {journal} {\bibinfo  {journal} {npj Quantum Materials}\ }\textbf
  {\bibinfo {volume} {2}},\ \bibinfo {pages} {24} (\bibinfo {year}
  {2017})}\BibitemShut {NoStop}%
\bibitem [{\citenamefont {Vafek}\ and\ \citenamefont
  {Vishwanath}(2014)}]{Vafek2014}%
  \BibitemOpen
  \bibfield  {author} {\bibinfo {author} {\bibfnamefont {O.}~\bibnamefont
  {Vafek}}\ and\ \bibinfo {author} {\bibfnamefont {A.}~\bibnamefont
  {Vishwanath}},\ }\href {\doibase 10.1146/annurev-conmatphys-031113-133841}
  {\bibfield  {journal} {\bibinfo  {journal} {Annual Review of Condensed Matter
  Physics}\ }\textbf {\bibinfo {volume} {5}},\ \bibinfo {pages} {83} (\bibinfo
  {year} {2014})},\ \Eprint
  {http://arxiv.org/abs/https://doi.org/10.1146/annurev-conmatphys-031113-133841}
  {https://doi.org/10.1146/annurev-conmatphys-031113-133841} \BibitemShut
  {NoStop}%
\bibitem [{\citenamefont {Murakami}\ and\ \citenamefont
  {Kuga}(2008)}]{Murakami2008}%
  \BibitemOpen
  \bibfield  {author} {\bibinfo {author} {\bibfnamefont {S.}~\bibnamefont
  {Murakami}}\ and\ \bibinfo {author} {\bibfnamefont {S.-i.}\ \bibnamefont
  {Kuga}},\ }\href {\doibase 10.1103/PhysRevB.78.165313} {\bibfield  {journal}
  {\bibinfo  {journal} {Phys. Rev. B}\ }\textbf {\bibinfo {volume} {78}},\
  \bibinfo {pages} {165313} (\bibinfo {year} {2008})}\BibitemShut {NoStop}%
\bibitem [{\citenamefont {Murakami}(2007)}]{Murakami2007}%
  \BibitemOpen
  \bibfield  {author} {\bibinfo {author} {\bibfnamefont {S.}~\bibnamefont
  {Murakami}},\ }\href {http://stacks.iop.org/1367-2630/9/i=9/a=356} {\bibfield
   {journal} {\bibinfo  {journal} {New Journal of Physics}\ }\textbf {\bibinfo
  {volume} {9}},\ \bibinfo {pages} {356} (\bibinfo {year} {2007})}\BibitemShut
  {NoStop}%
\bibitem [{\citenamefont {Hasan}\ and\ \citenamefont {Kane}(2010)}]{Hasan2010}%
  \BibitemOpen
  \bibfield  {author} {\bibinfo {author} {\bibfnamefont {M.~Z.}\ \bibnamefont
  {Hasan}}\ and\ \bibinfo {author} {\bibfnamefont {C.~L.}\ \bibnamefont
  {Kane}},\ }\href {\doibase 10.1103/RevModPhys.82.3045} {\bibfield  {journal}
  {\bibinfo  {journal} {Rev. Mod. Phys.}\ }\textbf {\bibinfo {volume} {82}},\
  \bibinfo {pages} {3045} (\bibinfo {year} {2010})}\BibitemShut {NoStop}%
\bibitem [{\citenamefont {Bernevig}\ and\ \citenamefont
  {Hughes}(2013)}]{Bernevig2013}%
  \BibitemOpen
  \bibfield  {author} {\bibinfo {author} {\bibfnamefont {B.~A.}\ \bibnamefont
  {Bernevig}}\ and\ \bibinfo {author} {\bibfnamefont {T.~L.}\ \bibnamefont
  {Hughes}},\ }\href@noop {} {\emph {\bibinfo {title} {Topological insulators
  and topological superconductors}}}\ (\bibinfo  {publisher} {Princeton
  University Press},\ \bibinfo {year} {2013})\BibitemShut {NoStop}%
\bibitem [{\citenamefont {Fradkin}(2013)}]{Fradkin2013}%
  \BibitemOpen
  \bibfield  {author} {\bibinfo {author} {\bibfnamefont {E.}~\bibnamefont
  {Fradkin}},\ }\href@noop {} {\emph {\bibinfo {title} {Field theories of
  condensed matter physics}}}\ (\bibinfo  {publisher} {Cambridge University
  Press},\ \bibinfo {year} {2013})\BibitemShut {NoStop}%
\bibitem [{\citenamefont {Zak}(1989)}]{Zak1989}%
  \BibitemOpen
  \bibfield  {author} {\bibinfo {author} {\bibfnamefont {J.}~\bibnamefont
  {Zak}},\ }\href {\doibase 10.1103/PhysRevLett.62.2747} {\bibfield  {journal}
  {\bibinfo  {journal} {Phys. Rev. Lett.}\ }\textbf {\bibinfo {volume} {62}},\
  \bibinfo {pages} {2747} (\bibinfo {year} {1989})}\BibitemShut {NoStop}%
\bibitem [{\citenamefont {Sato}\ and\ \citenamefont {Ando}(2017)}]{Sato2017}%
  \BibitemOpen
  \bibfield  {author} {\bibinfo {author} {\bibfnamefont {M.}~\bibnamefont
  {Sato}}\ and\ \bibinfo {author} {\bibfnamefont {Y.}~\bibnamefont {Ando}},\
  }\href@noop {} {\bibfield  {journal} {\bibinfo  {journal} {Reports on
  Progress in Physics}\ }\textbf {\bibinfo {volume} {80}},\ \bibinfo {pages}
  {076501} (\bibinfo {year} {2017})}\BibitemShut {NoStop}%
\bibitem [{\citenamefont {Fukui}\ \emph {et~al.}(2005)\citenamefont {Fukui},
  \citenamefont {Hatsugai},\ and\ \citenamefont {Suzuki}}]{Fukui2005}%
  \BibitemOpen
  \bibfield  {author} {\bibinfo {author} {\bibfnamefont {T.}~\bibnamefont
  {Fukui}}, \bibinfo {author} {\bibfnamefont {Y.}~\bibnamefont {Hatsugai}}, \
  and\ \bibinfo {author} {\bibfnamefont {H.}~\bibnamefont {Suzuki}},\ }\href
  {\doibase 10.1143/JPSJ.74.1674} {\bibfield  {journal} {\bibinfo  {journal}
  {Journal of the Physical Society of Japan}\ }\textbf {\bibinfo {volume}
  {74}},\ \bibinfo {pages} {1674} (\bibinfo {year} {2005})},\ \Eprint
  {http://arxiv.org/abs/0503172} {arXiv:0503172 [cond-mat]} \BibitemShut
  {NoStop}%
\bibitem [{\citenamefont {Vorontsov}\ \emph {et~al.}(2010)\citenamefont
  {Vorontsov}, \citenamefont {Vavilov},\ and\ \citenamefont
  {Chubukov}}]{Vorontsov2010}%
  \BibitemOpen
  \bibfield  {author} {\bibinfo {author} {\bibfnamefont {A.~B.}\ \bibnamefont
  {Vorontsov}}, \bibinfo {author} {\bibfnamefont {M.~G.}\ \bibnamefont
  {Vavilov}}, \ and\ \bibinfo {author} {\bibfnamefont {A.~V.}\ \bibnamefont
  {Chubukov}},\ }\href {\doibase 10.1103/PhysRevB.81.174538} {\bibfield
  {journal} {\bibinfo  {journal} {Phys. Rev. B}\ }\textbf {\bibinfo {volume}
  {81}},\ \bibinfo {pages} {174538} (\bibinfo {year} {2010})}\BibitemShut
  {NoStop}%
\bibitem [{\citenamefont {Heinzner}\ \emph {et~al.}(2005)\citenamefont
  {Heinzner}, \citenamefont {Huckleberry},\ and\ \citenamefont
  {Zirnbauer}}]{Heinzner2005}%
  \BibitemOpen
  \bibfield  {author} {\bibinfo {author} {\bibfnamefont {P.}~\bibnamefont
  {Heinzner}}, \bibinfo {author} {\bibfnamefont {A.}~\bibnamefont
  {Huckleberry}}, \ and\ \bibinfo {author} {\bibfnamefont {M.}~\bibnamefont
  {Zirnbauer}},\ }\href {\doibase 10.1007/s00220-005-1330-9} {\bibfield
  {journal} {\bibinfo  {journal} {Communications in Mathematical Physics}\
  }\textbf {\bibinfo {volume} {257}},\ \bibinfo {pages} {725} (\bibinfo {year}
  {2005})}\BibitemShut {NoStop}%
\end{thebibliography}%

\end{document}